\documentclass[prd,aps,showpacs,epsf,floats,10pt]{revtex4}%
\usepackage{amssymb}
\usepackage{amsfonts}
\usepackage{amsmath}
\usepackage{graphicx}%
\providecommand{\U}[1]{\protect\rule{.1in}{.1in}}

\begin{document}
\title{\textbf{Quantum Groverian Geodesic Paths with Gravitational and Thermal
Analogies}}
\author{\textbf{Carlo Cafaro}$^{1}$,\textbf{\ Domenico Felice}$^{2,1}$, and
\textbf{Paul M. Alsing}$^{3}$}
\affiliation{$^{1}$SUNY Polytechnic Institute, 12203 Albany, New York, USA}
\affiliation{$^{2}$I. S. I. S. \textquotedblleft A. Volta\textquotedblright, 81031 Aversa, Italy}
\affiliation{$^{3}$Air Force Research Laboratory, Information Directorate, 13441 Rome, New
York, USA}

\begin{abstract}
We present a unifying variational calculus derivation of Groverian geodesics
for both quantum state vectors and quantum probability amplitudes. In the
first case, we show that horizontal affinely parametrized geodesic paths on
the {Hilbert space of normalized vectors} emerge from the minimization of the
length specified by the Fubini-Study metric {on the manifold of Hilbert space
rays}. In the second case, we demonstrate that geodesic paths for probability
amplitudes arise by minimizing the length expressed in terms of the Fisher
information. In both derivations, we find that geodesic equations are
described by simple harmonic oscillators (SHOs). However, while in the first
derivation the frequency of oscillations is proportional to the (constant)
energy dispersion $\Delta E$ of the Hamiltonian system, in the second
derivation the frequency of oscillations is proportional to the square-root
$\sqrt{\mathcal{F}}$ of the (constant) Fisher information. Interestingly, by
setting these two frequencies equal to each other, we recover the well-known
Anandan-Aharonov relation linking the squared speed of evolution of an
Hamiltonian system with its energy dispersion. Finally, upon transitioning
away from the quantum setting, we discuss the universality of the emergence of
geodesic motion of SHO type in the presence of conserved quantities by
analyzing two specific phenomena of gravitational and thermodynamical origin, respectively.

\end{abstract}

\maketitle

\bigskip\pagebreak

\section{Introduction}

The concept of Fisher information plays a key role in both physics
\cite{frieden98} and information theory \cite{cover06}. {Methods of Fisher
information have been widely employed for both classical and quantum physical
systems \cite{Felice18}.} The increasing importance of the concept of Fisher
information in both statistical physics and quantum computing was recently
pointed out in \cite{carlopre}. In statistical physics for instance, the
application of Fisher information in the kinetic theory of gases is specified
by its decrease along the solutions of the Boltzmann equation for Maxwellian
molecules in the two-dimensional case \cite{villani00}. In quantum physics,
for example, the output state in Grover's quantum search algorithm follows a
geodesic path emerging from the Fubini-Study metric on the manifold of
Hilbert-space rays \cite{alvarez00,wadati01,carlophysica}.

In Ref. \cite{carlopre}, the authors presented an information geometric
characterization of the oscillatory or monotonic behavior of statistically
parametrized squared probability amplitudes originating from special
functional forms of the Fisher information function: constant, exponential
decay, and power-law decay. Furthermore, for each case, the authors computed
both the computational speed and the availability loss of the corresponding
physical processes by exploiting a convenient Riemannian geometrization of
useful thermodynamical concepts. In Ref. \cite{carlopre}, the authors also
commented on the possibility of using the proposed methods of information
geometry to help identify a suitable trade-off between speed and thermodynamic
efficiency in quantum search algorithms. The authors remarked that a deeper
understanding of the connection between the Fisher information and the
schedule of the quantum algorithm remained to be discovered in order to
provide a rigorous mapping between our information geometric analysis and the
Hamiltonian formulation of the quantum search problem. In particular, it
remained an open problem to describe in an exact quantitative
manner\textbf{\ }how the speed at which the search Hamiltonian can drive the
system toward the target state is related to both the functional forms of the
schedule and the Fisher information. Fortunately, the comprehension of the
connection between the Fisher information and the schedule of a quantum
algorithm has been highly enhanced in Refs. \cite{carloIJQI,carloPS,gassner20}%
. More specifically, a detailed investigation concerning the physical
connection between quantum search Hamiltonians and exactly solvable
time-dependent two-level quantum systems was presented in Ref.
\cite{carloIJQI}. In the work showed in Ref. \cite{carloIJQI}, the transition
probabilities from a source state to a target state in a number of physical
scenarios specified by a spin-$1/2$ particle immersed in an external
time-dependent magnetic field were computed in an exact analytical manner.
Both the periodic oscillatory as well as the monotonic temporal behaviors of
such transition probabilities were analyzed and their analogy with
characteristic features of Grover-like and fixed-point quantum search
algorithms were explored, respectively. Finally, the connection between the
schedule of a search algorithm in both adiabatic and nonadiabatic quantum
mechanical evolutions, and the control fields in a time-dependent driving
Hamiltonian was explored. In Ref. \cite{carlopre20}, motivated by the lack of
any comparative thermodynamical analysis of quantum search algorithms and
building on the previous works presented in Refs.
\cite{carlopre,carloIJQI,carloPS}, the authors borrowed the idea of Riemannian
geometrization of the concepts of efficiency and speed within both quantum and
thermodynamical settings in order to provide a theoretical perspective on the
trade-off between speed and efficiency in terms of minimum entropy production
paths emerging from quantum mechanical evolutions. Specifically, they
presented an information geometric analysis of entropic speeds and entropy
production rates in geodesic evolution on statistical manifolds of
parametrized quantum states arising as outputs of \textrm{su}$\left(  2\text{;
}%
\mathbb{C}
\right)  $ Hamiltonian models mimicking different types of continuous-time
quantum search evolutions. Interestingly, building upon the work presented in
Ref. \cite{carlopre20}, the same authors have recently presented an
information geometric analysis of off-resonance effects on classes of exactly
solvable generalized semi-classical Rabi systems in Ref. \cite{quantum20}.

After a serious reconsideration of the underlying geometrical structure of our
previously mentioned works in Refs.
\cite{carlopre,carloIJQI,carloPS,gassner20,carlopre20,quantum20,
carlophysica}, we have arrived at the conclusion that it is the effective and
serious exploitation of the link between quantum search algorithms and
geodesics in the complex projective Hilbert space that serves as the essential
ingredient giving rise to an increasing number of intriguing interdisciplinary
investigations in the literature connecting concepts from information
geometry, quantum computing, and thermodynamics. For that reason, we propose
to reconsider in this article the mathematical derivation of such an important
mathematical link that provides an ever increasing number of penetrating
physical insights.

In this article, we provide a unifying variational calculus computation of
Groverian geodesics for both quantum state vectors and quantum probability
amplitudes. In the first case, following the works in Refs.
\cite{mukunda93,pati94,pati95}, we verify that horizontal affinely
parametrized geodesic paths on {the Hilbert space of normalized vectors can be
obtained from the minimization of the length specified by the Fubini-Study
metric on the manifold of Hilbert space rays}. In the second case, inspired by
the works in Refs. \cite{alvarez00,mancioA,mancioB} , we show that geodesic
paths for probability amplitudes emerge by minimizing the length specified by
means of the {classical} Fisher information $\mathcal{F}$. For the sake of
completeness, we point out that the {fully quantum scenario arises when phase
factors are }also{ considered. In such a case, quantum geodesics show a
behavior different from that of geodesics emerging from the classical Fisher
information \cite{Ciaglia18}.} For both derivations, we observe that geodesic
equations describe simple harmonic oscillators (SHOs) with distinct
frequencies. Specifically, in the first derivation the frequency of
oscillations is proportional to the (constant) energy dispersion $\Delta E$ of
the Hamiltonian system. In the second derivation, by contrast, the frequency
of oscillations is proportional to the square-root $\sqrt{\mathcal{F}}$ of the
(constant) Fisher information. As a pleasant result, we find that
upon\textbf{\ }equating these two frequencies, we retrieve the well-known
Anandan-Aharonov relation (for instance, see Ref. \cite{anandan90}) linking
the squared speed $v_{\mathrm{H}}^{2}$ of evolution of an Hamiltonian
(\textrm{H}) system with its energy dispersion $\Delta E$ together with the
neat link between speed of evolution $v_{\mathrm{H}}$ and Fisher information
$\mathcal{F}$,\textbf{\ }namely%
\begin{equation}
\frac{\Delta E^{2}}{\hslash^{2}}=v_{\mathrm{H}}^{2}=\frac{\mathcal{F}}%
{4}\text{,} \label{UNO}%
\end{equation}
with $\hslash\overset{\text{def}}{=}h/\left(  2\pi\right)  $ being the reduced
Planck constant. Finally, moving away from the quantum mechanical setting, we
elaborate on the universality of the emergence of geodesic motion of SHO type
in the presence of conserved quantities by discussing two specific phenomena
of gravitational and thermodynamical flavor, respectively.

The layout of the remainder of this article is as follows. In Section II, we
briefly summarize the essential features of Grover's quantum search algorithm
\cite{grover}. In particular, we focus on its discrete output quantum state
after $k$-iterations and its continuous version in terms of a parametric
quantum state vector. In Section III, we discuss two distinct variational
calculus derivations yielding Grover-like (Groverian) geodesics. In the first
computation, we find the most general parametrization of an horizontal
geodesic path in the space of unit rays assuming that the energy dispersion
$\Delta E$ of the systems is constant. In the second computation, using the
fact that the Fisher information $\mathcal{F}$ for a Groverian probability
path is constant, we derive the most general parametrization for the quantum
probability amplitudes specifying the quantum state vector tracing the
geodesic path. Both geodesic equations describe a simple harmonic oscillator
(SHO). Moreover, upon equating the frequencies of the two oscillators, we
recover (as pointed out earlier) the Anandan-Aharonov speed-energy relation.
In Section IV, upon transitioning away from quantum physics, we move to
general relativity and thermodynamics. We focus on two physical scenarios
where the geodesic motion becomes of SHO type under specific physical
conditions. In the gravitational \cite{zatzkis59} and thermal
\cite{ruchhardt29} settings, the frequencies of the SHOs are expressed in
terms of the mass density $\rho$ of an ideal liquid and the adiabatic
coefficient $\gamma$ of an ideal gas, respectively. Our discussion and final
remarks appear in Section V. Finally, some more technical details can be found
in Appendices A, B, and C.

\section{Grover's quantum search algorithm}

In this Section we present a brief summary of Grover's quantum search protocol
\cite{grover}. The searching problem addressed by Grover's algorithm may be
re-stated as follows: assume we seek to retrieve a specific database entry
subject to some previously specified condition, provided the entry in question
belongs to an unsorted database (i.e. oracle) containing $N\overset
{\text{def}}{=}$ $2^{n}$ elements with $N$ denoting the dimensionality of the
complex Hilbert space $\mathcal{H}$ with $n$-qubit quantum states. One step is
required to determine whether the entry that has been examined is the one
satisfying the given condition. We assume further that finding the selected
entry is not aided by sorting of the database. In a situation such as that
described above, the maximally efficient classical algorithm capable of
implementing the search scheme necessarily requires examination of database
elements one at a time. Thus, if a classical computer is utilized to carry out
the searching protocol, then the oracle must be queried on average $N/2$ times
(i.e. $\mathcal{O}\left(  N\right)  $ classical steps). By utilizing the same
hardware as in the classical case however, but requiring the input and output
be in a superposition of states, Grover developed a quantum algorithm capable
of implementing this searching problem in approximately $\frac{\pi}{4}\sqrt
{N}$ steps (i.e. $\mathcal{O}\left(  \sqrt{N}\right)  $ quantum mechanical
steps) \cite{grover}.

When considering the case of $n$-qubits quantum states, the construction of
Grover's search algorithm can be described as follows \cite{nielsen}. The
initialization\ (i.e. step-$0$) of Grover's algorithm commences with an
application of the Hadamard transform for the purpose of constructing an
initial state with uniform amplitude. This initial state is comprised of an
equal superposition of all orthonormal computational basis states $\left\{
\left\vert w\right\rangle \right\}  $\ belonging to the $N$-dimensional
Hilbert space,%
\begin{equation}
\left\vert s\right\rangle \overset{\text{def}}{=}\frac{1}{\sqrt{N}}\sum
_{w=1}^{N}\left\vert w\right\rangle =\sin\left(  \frac{\varphi}{2}\right)
\left\vert \bar{w}\right\rangle +\cos\left(  \frac{\varphi}{2}\right)
\left\vert w_{\perp}\right\rangle \text{.} \label{input state}%
\end{equation}
The state $\left\vert w_{\perp}\right\rangle $\ in Eq. (\ref{input state}) is
defined as,%
\begin{equation}
\left\vert w_{\perp}\right\rangle \overset{\text{def}}{=}\sqrt{\frac{1}{N-1}%
}\sum_{w\neq\bar{w}}\left\vert w\right\rangle \label{sin}%
\end{equation}
while the angle $\varphi$ quantifies the overlap between the (source) state
$\left\vert s\right\rangle $ and the (target) state $\left\vert \bar
{w}\right\rangle $ and is given by%
\begin{equation}
\sin\left(  \frac{\varphi}{2}\right)  \overset{\text{def}}{=}\frac{1}{\sqrt
{N}}=\left\langle \bar{w}|s\right\rangle \text{.}%
\end{equation}
For example, given an initial input, a single iteration of Grover's algorithm
induces a rotation by angle $\varphi$ in the two-dimensional space spanned by
states $\left\vert w_{\perp}\right\rangle $ and $\left\vert \bar
{w}\right\rangle $. After $k$-iterations, the algorithm arrives at the state
$\left\vert \psi_{\text{Grover}}\left(  k\right)  \right\rangle $ written as%
\begin{equation}
\left\vert \psi_{\text{Grover}}\left(  k\right)  \right\rangle \overset
{\text{def}}{=}G^{k}\left\vert s\right\rangle =\sin\left[  \left(  k+\frac
{1}{2}\right)  \varphi\right]  \left\vert \bar{w}\right\rangle +\cos\left[
\left(  k+\frac{1}{2}\right)  \varphi\right]  \left\vert w_{\perp
}\right\rangle \text{,} \label{output}%
\end{equation}
with $G$ denoting the so-called Grover iterate \cite{nielsen}. In the limit
where $N\gg1$, the number of iterations $\bar{k}$ for which $\left\vert
\psi_{\text{Grover}}\left(  \bar{k}\right)  \right\rangle $ is identical to
the target state $\left\vert \bar{w}\right\rangle $ (i.e. when the algorithm
achieves success probability equal to one) is approximated by%
\begin{equation}
\bar{k}\overset{N\gg1}{\simeq}\frac{\pi}{4}\sqrt{N}\text{.} \label{finalino}%
\end{equation}
Equation (\ref{finalino}) is obtained by requiring $\left(  \bar{k}+\frac
{1}{2}\right)  \varphi=\frac{\pi}{2}$ and by recognizing that when $N\gg1$,
Eq. (\ref{sin}) implies $\varphi\simeq2/\sqrt{N}$. We observe that Grover's
search algorithm evolves with discrete $k$. However, the temporal interval
between two consecutive discrete steps $\Delta t\overset{\text{def}}{=}%
t_{k+1}-t_{k}=\varphi/2\approx N^{-1/2}$ becomes infinitesimally small when
$N$ assumes sufficiently large values \cite{wadati01}. Thus, in this
particular limiting scenario, we can approximately identify $\left(
k+1/2\right)  \varphi$ with a continuous parameter $\theta$ in such a manner
that no skipping occurs along an hypothetical geodesic motion. Specifically,
the output state (\ref{output}) can be well approximated by a state vector
$\left\vert \psi_{\text{Grover}}\left(  \theta\right)  \right\rangle $ that
depends upon a continuous parameter $\theta$ in the limit where $N\gg1$,
namely%
\begin{equation}
\left\vert \psi_{\text{Grover}}\left(  \theta\right)  \right\rangle
\overset{\text{def}}{=}\sum_{m=1}^{N}\sqrt{p_{m}\left(  \theta\right)
}\left\vert m\right\rangle \text{,} \label{conti}%
\end{equation}
where,%
\begin{equation}
\left\langle m|m^{\prime}\right\rangle =\delta_{mm^{\prime}}\text{, }%
p_{1}\left(  \theta\right)  \overset{\text{def}}{=}\sin^{2}\theta\text{ and,
}p_{l}\left(  \theta\right)  \overset{\text{def}}{=}\frac{\cos^{2}\theta}%
{N-1}\text{ with }l\neq1\text{.} \label{probabilities}%
\end{equation}
Indeed, the $N$-dimensional vector of probability distributions $\vec
{p}\overset{\text{def}}{=}\left(  p_{1}\left(  \theta\right)  \text{, }%
p_{2}\left(  \theta\right)  \text{,..., }p_{N}\left(  \theta\right)  \right)
$ with $p_{j}\left(  \theta\right)  $ defined in (\ref{probabilities}) can be
interpreted as a path induced by Grover's search algorithm on some suitable
probability manifold. In the second part of the following Section, we
demonstrate that such a probability path is in fact a geodesic curve for which
the Fisher information action functional $\mathcal{S}\left[  p_{m}\left(
\theta\right)  \right]  \overset{\text{def}}{=}\frac{1}{2}\int\sqrt
{\mathcal{F}\left(  \theta\right)  }d\theta$ with the Fisher information
function $\mathcal{F}\left(  \theta\right)  $ defined as \cite{frieden98},
\begin{equation}
\mathcal{F}\left(  \theta\right)  \overset{\text{def}}{=}4\sum_{m=1}%
^{N}\left(  \frac{\partial\sqrt{p_{m}}}{\partial\theta}\right)  ^{2}\text{,}
\label{FI}%
\end{equation}
is extremized. For further details on Grover's quantum search algorithm and
Fisher's information, we refer to Ref. \cite{nielsen} and Ref.
\cite{frieden98}, respectively.

\section{Groverian geodesics}

In this Section, we present two distinct variational calculus derivations
yielding Groverian geodesics. In the first derivation, assuming that the
energy dispersion $\Delta E$ of the systems is constant, we obtain the most
general parametrization of an horizontal geodesic path in the space of unit
rays. In the second derivation, exploiting the fact that the Fisher
information $\mathcal{F}$ for a Groverian probability path is constant
\cite{alvarez00,mancioA,mancioB}, we obtain the most general parametrization
for the quantum probability amplitudes specifying the quantum state vector
tracing the geodesic path. We emphasize that technical details on geodesics in
the space of unit rays with special focus on horizontal affinely parameterized
geodesics appear in Appendix A.

\subsection{Geodesic paths for quantum state vectors}

Geodesic paths $\gamma_{\mathrm{geo}}\left(  s\right)  $ with $s_{1}\leq s\leq
s_{2}$ in the {Hilbert space rays} are those for which the length functional
$\mathcal{L}\left[  \gamma_{\mathrm{geo}}\right]  $ \cite{mukunda93},%
\begin{equation}
\mathcal{L}\left[  \gamma_{\mathrm{geo}}\right]  \overset{\text{def}}{=}%
\int_{s_{1}}^{s_{2}}\sqrt{ds_{\text{FS}}^{2}}=\int_{s_{1}}^{s_{2}}\left\langle
u_{\perp}\left(  s\right)  |u_{\perp}\left(  s\right)  \right\rangle
^{1/2}ds\text{,}%
\end{equation}
is stationary. Note that $ds_{\text{FS}}^{2}$ is the Fubini-Study metric on
the projective Hilbert space $\mathcal{P}\left(  \mathcal{H}\right)  \simeq$ $%
\mathbb{C}
\mathcal{P}^{N-1}$ with $\mathcal{H}\simeq%
\mathbb{C}
^{N}$ defined as \cite{provost80,braunstein94},%
\begin{equation}
ds_{\text{FS}}^{2}\overset{\text{def}}{=}\left\langle u_{\perp}\left(
s\right)  |u_{\perp}\left(  s\right)  \right\rangle ds^{2}=\left\langle
\dot{\psi}_{\perp}\left(  s\right)  |\dot{\psi}_{\perp}\left(  s\right)
\right\rangle ds^{2}\text{,}\label{FSyes}%
\end{equation}
where $\left\vert u_{\perp}\left(  s\right)  \right\rangle \overset
{\text{def}}{=}\left\vert \dot{\psi}_{\perp}\left(  s\right)  \right\rangle $
with $\dot{\psi}_{\perp}\left(  s\right)  \overset{\text{def}}{=}\partial
_{s}\psi_{\perp}$ and $\partial_{s}\overset{\text{def}}{=}\partial/\partial
s$. Furthermore, $\left\vert d\psi_{\perp}\right\rangle \overset{\text{def}%
}{=}\left\vert d\psi\right\rangle -\left\langle \psi|d\psi\right\rangle
\left\vert \psi\right\rangle $ is the projection of $\left\vert d\psi
\right\rangle $ orthogonal to $\left\vert \psi\right\rangle $. We remark that
$\left\vert d\psi\right\rangle \overset{\text{def}}{=}\left\vert \psi^{\prime
}\right\rangle -\left\vert \psi\right\rangle $ is the difference between the
two neighboring normalized pure states $\left\vert \psi\right\rangle $ and
$\left\vert \psi^{\prime}\right\rangle $. Stationarity of the length
functional requires that $\delta\mathcal{L}\left[  \gamma_{\mathrm{geo}%
}\right]  =0$ for arbitrary variations $\left\vert \delta\psi\right\rangle $,
subject only to the constraint equation $\operatorname{Re}\left[  \left\langle
\delta\psi|\psi\right\rangle \right]  =0$. By carrying out a variational
calculation, it is possible to show that when the uncertainty $\Delta E\left(
t\right)  $ in the energy,%
\begin{equation}
\Delta E\left(  t\right)  \overset{\text{def}}{=}\left[  \left\langle
\psi\left(  t\right)  |\mathrm{H}^{2}\left(  t\right)  |\psi\left(  t\right)
\right\rangle -\left\langle \psi\left(  t\right)  |\mathrm{H}\left(  t\right)
|\psi\left(  t\right)  \right\rangle ^{2}\right]  ^{1/2}\text{,}%
\end{equation}
does not depend on time, the most general equation in horizontal and affinely
parametrized form ($\left\vert \psi_{\mathrm{h}}\left(  s\right)
\right\rangle $) is given by a simple harmonic oscillator equation,%
\begin{equation}
\frac{d^{2}}{ds^{2}}\left\vert \psi_{\mathrm{h}}\left(  s\right)
\right\rangle +v_{\mathrm{H}}^{2}\left\vert \psi_{\mathrm{h}}\left(  s\right)
\right\rangle =0\text{.}\label{GE1}%
\end{equation}
It can be shown that Eq. (\ref{GE1}) emerges from a variational calculus
computation where one seeks to extremize the length functional $\mathcal{L}%
\left[  \gamma_{\mathrm{geo}}\right]  $ and, at the same time, exploits both
the gauge invariance and the reparametrization invariance of such a
functional. Then, $\left\vert \psi_{\mathrm{h}}\left(  s\right)  \right\rangle
$ in Eq. (\ref{GE1}) denotes the most general geodesic in horizontal and
affinely parametrized form \cite{mukunda93}. In particular, while gauge
invariance leads to the so-called horizontality condition, reparametrization
invariance motivates affine parametrizations. An explicit computation with all
technical details yielding Eq. (\ref{GE1}) together with its most general
solution $\left\vert \psi_{\mathrm{h}}\left(  s\right)  \right\rangle $
appears in Appendix A. The generally $s$-dependent quantity $v_{\mathrm{H}%
}\left(  s\right)  $ in Eq. (\ref{GE1}),%
\begin{equation}
v_{\mathrm{H}}\left(  s\right)  \overset{\text{def}}{=}\left\langle \dot{\psi
}_{\mathrm{h}}\left(  s\right)  |\dot{\psi}_{\mathrm{h}}\left(  s\right)
\right\rangle ^{1/2}=\frac{\Delta E\left(  s\right)  }{\hslash}\text{,}%
\end{equation}
denotes the speed of transportation of the parallel transported horizontal
state vector $\left\vert \psi_{\mathrm{h}}\left(  s\right)  \right\rangle $
{in the Hilbert space of normalized vectors} satisfying the parallel transport
rule, $\left\langle \psi_{\mathrm{h}}\left(  s\right)  |\dot{\psi}%
_{\mathrm{h}}\left(  s\right)  \right\rangle =0$. For the sake of
completeness, we point out that the state vector $\left\vert \psi_{\mathrm{h}%
}\left(  s\right)  \right\rangle $ is connected to the dynamical state vector
$\left\vert \psi\left(  s\right)  \right\rangle $ satisfying the\textbf{
}Schr\"{o}dinger evolution equation by the relation \cite{pati94,pati95},%
\begin{equation}
\left\vert \psi_{\mathrm{h}}\left(  s\right)  \right\rangle \overset
{\text{def}}{=}\exp\left[  \left(  \frac{i}{\hslash}\right)  \int_{0}%
^{s}\left\langle \psi\left(  s^{\prime}\right)  |\mathrm{H}\left(  s^{\prime
}\right)  |\psi\left(  s^{\prime}\right)  \right\rangle ds^{\prime}\right]
\left\vert \psi\left(  s\right)  \right\rangle \text{.}%
\end{equation}
Finally, assuming that $\left\langle \psi_{\mathrm{h}}\left(  0\right)
|\psi_{\mathrm{h}}\left(  0\right)  \right\rangle =1$, $\left\langle
\psi_{\mathrm{h}}\left(  0\right)  |\dot{\psi}_{\mathrm{h}}\left(  0\right)
\right\rangle =0$, and $\left\langle \dot{\psi}_{\mathrm{h}}\left(  0\right)
|\dot{\psi}_{\mathrm{h}}\left(  0\right)  \right\rangle =v_{\mathrm{H}}^{2}=$
\textrm{constant}, the general solution to Eq. (\ref{GE1}) becomes%
\begin{equation}
\left\vert \psi_{\mathrm{h}}\left(  s\right)  \right\rangle =\cos\left(
v_{\mathrm{H}}s\right)  \left\vert \psi_{\mathrm{h}}\left(  0\right)
\right\rangle +\frac{\sin\left(  v_{\mathrm{H}}s\right)  }{v_{\mathrm{H}}%
}\left\vert \dot{\psi}_{\mathrm{h}}\left(  0\right)  \right\rangle
\text{,}\label{GT1}%
\end{equation}
where $v_{\mathrm{H}}$ in Eq. (\ref{GT1}) is defined as,%
\begin{equation}
v_{\mathrm{H}}\overset{\text{def}}{=}\frac{\Delta E}{\hslash}%
=\mathrm{constant}\text{.}\label{F1}%
\end{equation}
For the sake of completeness, we remark that the information encoded into the
three conditions right before Eq. (\ref{GT1}) determines completely the curve
$s\mapsto\left\vert \psi_{\mathrm{h}}\left(  s\right)  \right\rangle $. In
particular, these conditions specify that the curve is traced by starting on
the sphere ($\left\langle \psi_{\mathrm{h}}\left(  0\right)  |\psi
_{\mathrm{h}}\left(  0\right)  \right\rangle =1$) with constant velocity
$v_{\mathrm{H}}$ ($\left\langle \dot{\psi}_{\mathrm{h}}\left(  0\right)
|\dot{\psi}_{\mathrm{h}}\left(  0\right)  \right\rangle =v_{\mathrm{H}}^{2}=$
constant) in a direction tangent to the sphere ($\left\langle \psi
_{\mathrm{h}}\left(  0\right)  |\dot{\psi}_{\mathrm{h}}\left(  0\right)
\right\rangle =0$). From Eq. (\ref{GT1}), we conclude that the horizontal
geodesic can be geometrically interpreted as a real two-dimensional rotation
on the plane spanned by the state vectors $\left\vert \psi_{\mathrm{h}}\left(
0\right)  \right\rangle $ and $\left\vert \dot{\psi}_{\mathrm{h}}\left(
0\right)  \right\rangle $. It therefore follows that we can view the
transition probability $\mathrm{P}_{\left\vert \psi_{\mathrm{h}}\left(
0\right)  \right\rangle \rightarrow\left\vert \psi_{\mathrm{h}}\left(
s\right)  \right\rangle }\left(  s\right)  $ from $\left\vert \psi
_{\mathrm{h}}\left(  0\right)  \right\rangle $ to $\left\vert \psi
_{\mathrm{h}}\left(  s\right)  \right\rangle $,%
\begin{equation}
\mathrm{P}_{\left\vert \psi_{\mathrm{h}}\left(  0\right)  \right\rangle
\rightarrow\left\vert \psi_{\mathrm{h}}\left(  s\right)  \right\rangle
}\left(  s\right)  \overset{\text{def}}{=}\left\vert \left\langle
\psi_{\mathrm{h}}\left(  s\right)  |\psi_{\mathrm{h}}\left(  0\right)
\right\rangle \right\vert ^{2}\text{,}%
\end{equation}
in terms of the distance $s$ along the geodesic joining $\left\vert
\psi_{\mathrm{h}}\left(  0\right)  \right\rangle $ and $\left\vert
\psi_{\mathrm{h}}\left(  s\right)  \right\rangle $. The transition from the
digital (discrete) to the analog (continuous time) evolution of Grover's
quantum search algorithm occurs in the limit of $N\gg1$ where the interval of
skip $\Delta k/k=\left[  \left(  k+1/2\right)  \varphi-\varphi/2\right]
/k=\varphi\simeq2/\sqrt{N}$ between two consecutive steps becomes
infinitesimally small. In such a limit, the Riemannian geometric formulation
of quantum mechanics suggests that Grover's dynamics yields the shortest paths
in $\mathcal{%
\mathbb{C}
P}^{N-1}$. Indeed, upon\ setting $v_{\mathrm{H}}=1$, Groverian geodesics
$\left\vert \psi_{\text{Grover}}\left(  s\right)  \right\rangle \overset
{\text{def}}{=}\cos\left(  s\right)  \left\vert w_{\perp}\right\rangle
+\sin\left(  s\right)  \left\vert w\right\rangle $ can be formally obtained
from Eq. (\ref{GT1}) by taking $\left\vert \psi_{\mathrm{h}}\left(  0\right)
\right\rangle =\left\vert w_{\perp}\right\rangle $, $\left\vert \dot{\psi
}_{\mathrm{h}}\left(  0\right)  \right\rangle =\left\vert w\right\rangle $,
and $s=\left(  k+1/2\right)  \varphi$.

In the following subsection, we obtain the most general parametrization for
the quantum probability amplitudes specifying the quantum state vector tracing
a quantum Groverian geodesic path.

\subsection{Geodesic paths for quantum probability amplitudes}

The natural notion of distance between two neighboring pure physical states,
the so-called angle in Hilbert space, is specified by the Fubini-Study metric.
Except for a constant factor, this is the only Riemannian metric on the set of
rays in Hilbert space which is invariant under all unitary transformations.
The probabilistic nature of quantum mechanics in its geometric formulation can
be made more transparent by considering the equivalence between the angle in
Hilbert space and the statistical distance introduced by Wootters to
statistically distinguish between two different rays in the same Hilbert space
\cite{wootters81}. Within this statistical geometric framework of quantum
mechanics, to a greater distance between two neighboring pure states
{$\left\vert \psi\right\rangle +\left\vert d\psi\right\rangle $ and
$\left\vert \psi\right\rangle $ \cite{braunstein94},%
\begin{equation}
\left\vert \widetilde{\psi}\right\rangle \overset{\text{def}}{=}\left\vert
\psi\right\rangle +\left\vert d\psi\right\rangle {=}\sum_{m=1}^{N}\sqrt
{p_{m}+dp_{m}}e^{i\left(  \phi_{m}+d\phi_{m}\right)  }\left\vert
m\right\rangle \text{, and }\left\vert \psi\right\rangle \overset{\text{def}%
}{=}\sum_{m=1}^{N}\sqrt{p_{m}}e^{i\phi_{m}}\left\vert m\right\rangle
\text{,}\label{para}%
\end{equation}
there corresponds a higher degree of distinguishability of the two states.}
{Here} $\left\{  \left\vert m\right\rangle \right\}  $ $m\in\left\{
1\text{,..., }N\right\}  $ is an orthonormal basis of the $N$-dimensional
complex Hilbert space $\mathcal{H}$. By using Eq. (\ref{para}) {and the
Fubini-Study metric $ds_{\text{FS}}^{2}$ in Eq. (\ref{FSyes}) }reads as
follows \cite{braunstein94},
\begin{equation}
ds_{\text{FS}}^{2}\overset{\text{def}}{=}\left\langle d\psi_{\bot}|d\psi
_{\bot}\right\rangle =1-\left\vert \left\langle \widetilde{\psi}%
|\psi\right\rangle \right\vert ^{2}\text{,}%
\end{equation}
with $\left\vert d\psi_{\bot}\right\rangle \overset{\text{def}}{=}\left\vert
d\psi\right\rangle -\left\vert \psi\right\rangle \left\langle \psi
|d\psi\right\rangle $. {Now consider that both, $\left\vert \widetilde{\psi
}\right\rangle $ and $\left\vert {\psi}\right\rangle $, are parametrized by a
family $M=\{\mathbf{\theta}=(\theta^{1},\ldots,\theta^{n})\}\subset
\mathbb{R}^{n}$, i.e. $p_{m}\equiv p_{m}(\theta)$ and $\phi_{m}\equiv\phi
_{m}(\theta)$. Assuming that the parametrization is one-to-one, the
representation given in Eq. \eqref{para} provides an embedding of the family
$M$ into the Hilbert space of rays. Therefore, the Fubini-Study metric
\eqref{FSyes} can be pulled-back to $M$ to obtain the following expression
\cite{Ciaglia18}:
\begin{equation}
\mathcal{F}_{q}(\theta)=\frac{1}{4}\left\{  \sum_{m=1}^{N}\frac{(d{p}_{m}%
)^{2}}{p_{m}}+4\left[  \sum_{m=1}^{N}p_{m}(d{\phi}_{m})^{2}-\left(  \sum
_{m=1}^{N}p_{m}d{\phi}_{m}\right)  ^{2}\right]  \right\}  \text{,}%
\label{fessa}%
\end{equation}
which is the quantum version of the Fisher information \cite{Marmo}.}
{Obviously, if $M\subset\mathbb{R}$ the expression above becomes
\begin{equation}
\mathcal{F}_{q}(\theta)=\frac{1}{4}\left\{  \sum_{m=1}^{N}\frac{(\dot{p}%
_{m})^{2}}{p_{m}}+4\left[  \sum_{m=1}^{N}p_{m}(\dot{\phi}_{m})^{2}-\left(
\sum_{m=1}^{N}p_{m}\dot{\phi}_{m}\right)  ^{2}\right]  \right\}  d\theta
^{2}\text{,}\label{lineparam}%
\end{equation}
with $\dot{p}_{m}\overset{\text{def}}{=}dp_{m}/d\theta$ and $\dot{\phi}%
_{m}\overset{\text{def}}{=}d\phi_{m}/d\theta$}. We emphasize that it is always
possible to assume the variance of phase changes $\sigma_{\dot{\phi}}^{2}$,%
\begin{equation}
\sigma_{\dot{\phi}}^{2}\overset{\text{def}}{=}\sum_{m=1}^{N}p_{m}\dot{\phi
}_{m}^{2}-\left(  \sum_{m=1}^{N}p_{m}\dot{\phi}_{m}\right)  ^{2}%
\text{,}\label{variance1}%
\end{equation}
to be equal to zero upon selecting an appropriate choice of basis $\left\{
\left\vert m\right\rangle \right\}  $ \cite{braunstein94}. In what follows, we
assume to be reasoning with Eq. \eqref{lineparam} under such a working
condition (for further details on this particular matter, see Refs.
\cite{carlopre,quantum20}). Then, geodesic probability paths $\gamma
_{\mathrm{geo}}\left(  \theta\right)  $ with $\theta_{1}\leq\theta\leq
\theta_{2}$ in the probability space associated with the complex projective
Hilbert space $\mathcal{%
\mathbb{C}
P}^{N-1}$ are determined by\ minimization of the action $\mathcal{S}\left[
p_{m}\left(  \theta\right)  \right]  $,%
\begin{equation}
\mathcal{S}\left[  p_{m}\left(  \theta\right)  \right]  =\int\mathcal{L}%
\left(  \dot{p}_{m}\left(  \theta\right)  \text{, }p_{m}\left(  \theta\right)
\text{, }\theta\right)  d\theta\text{,}%
\end{equation}
where the Lagrangian-like quantity $\mathcal{L}\left(  \dot{p}_{m}\left(
\theta\right)  \text{, }p_{m}\left(  \theta\right)  \text{, }\theta\right)  $
is given by,%
\begin{equation}
\mathcal{L}\left(  \dot{p}_{m}\left(  \theta\right)  \text{, }p_{m}\left(
\theta\right)  \text{, }\theta\right)  \overset{\text{def}}{=}\frac{1}{2}%
\sqrt{\mathcal{F}\left(  \theta\right)  }=\frac{1}{2}\left[  \sum_{m=1}%
^{N}\frac{\dot{p}_{m}^{2}\left(  \theta\right)  }{p_{m}\left(  \theta\right)
}\right]  ^{\frac{1}{2}}\text{, }\label{lagrangian}%
\end{equation}
subject to the normalization constraint on the parametrized probability
distribution functions $p_{m}\left(  \theta\right)  $,%
\begin{equation}
\sum_{m=1}^{N}p_{m}\left(  \theta\right)  =1\text{.}\label{nc}%
\end{equation}
{Note that the Fisher information $\mathcal{F}$ entering the Eq. \eqref{nc} is
the classical one. This should come as no surprise since requiring
\eqref{variance1} amounts to neglect the phase factor from our analysis.} For
ease of analysis, we consider the change of variable $p_{m}\left(
\theta\right)  \rightarrow q_{m}^{2}\left(  \theta\right)  $ \cite{wootters81}%
. Then by making use of the method of Lagrange multipliers with the new
variable $q_{m}\left(  \theta\right)  $, we seek to minimize the new action
$\mathcal{S}^{\prime}\left[  q_{m}\left(  \theta\right)  \right]  $%
\begin{equation}
\mathcal{S}^{\prime}\left[  q_{m}\left(  \theta\right)  \right]
=\int\mathcal{L}^{\prime}\left(  \dot{q}_{m}\left(  \theta\right)  \text{,
}q_{m}\left(  \theta\right)  \text{, }\theta\right)  d\theta=\int\left\{
\left[  \sum_{m=1}^{N}\dot{q}_{m}^{2}\left(  \theta\right)  \right]
^{\frac{1}{2}}-\lambda_{\text{FS}}\left(  \sum_{m=1}^{N}q_{m}^{2}\left(
\theta\right)  -1\right)  \right\}  d\theta\text{.}\label{stay}%
\end{equation}
In Eq. (\ref{stay}), $\lambda_{\text{FS}}$\ is a Lagrange multiplier and
$\mathcal{L}^{\prime}\left(  \dot{q}_{m}\left(  \theta\right)  \text{, }%
q_{m}\left(  \theta\right)  \text{, }\theta\right)  $ acts as a new
Lagrangian-like quantity. The path that serves to minimize the action
$\mathcal{S}^{\prime}\left[  q_{m}\left(  \theta\right)  \right]  $ satisfies
the \textquotedblleft actuality constraint\textquotedblright,%
\begin{equation}
\frac{\delta\mathcal{S}^{\prime}\left[  q_{m}\left(  \theta\right)  \right]
}{\delta q_{m}\left(  \theta\right)  }=0\text{,}\label{minima}%
\end{equation}
leading to the following Euler-Lagrange (EL) equation in $q_{m}=q_{m}\left(
\theta\right)  $, namely%
\begin{equation}
\ddot{q}_{m}-\frac{\mathcal{\dot{L}}\left(  \theta\right)  }{\mathcal{L}%
\left(  \theta\right)  }\dot{q}_{m}+2\lambda_{\text{FS}}\mathcal{L}\left(
\theta\right)  q_{m}=0\text{,}\label{EL}%
\end{equation}
where $\mathcal{L}\left(  \theta\right)  \overset{\text{def}}{=}%
\mathcal{L}\left(  \dot{p}_{m}\left(  \theta\right)  \text{, }p_{m}\left(
\theta\right)  \text{, }\theta\right)  $ appears in Eq. (\ref{lagrangian})
while $\mathcal{\dot{L}}\left(  \theta\right)  \overset{\text{def}}%
{=}d\mathcal{L}/d\theta$. Given that $\mathcal{F}\left(  \theta\right)
=4\mathcal{L}^{2}\left(  \theta\right)  $, Eq. (\ref{EL}) can be recast as%
\begin{equation}
\ddot{q}_{m}\text{ }-\frac{1}{2}\frac{\mathcal{\dot{F}}\left(  \theta\right)
}{\mathcal{F}\left(  \theta\right)  }\dot{q}_{m}+\lambda_{\text{FS}%
}\mathcal{F}^{1/2}\left(  \theta\right)  q_{m}=0\text{.}\label{fsode}%
\end{equation}
Assuming $\mathcal{F}\left(  \theta\right)  =\mathcal{F}_{0}$ is constant, Eq.
(\ref{fsode}) reduces $\ddot{q}_{m}$ $+\lambda_{\text{FS}}\mathcal{F}%
^{1/2}\left(  \theta\right)  q_{m}=0$. The Lagrange multiplier $\lambda
_{\text{FS}}$\ is fixed by satisfying conservation of probability in Grover's
dynamics. Recalling that $\mathcal{F}_{0}=4$ in Grover's dynamics, this
constraint demands that $\lambda_{\text{FS}}$ satisfies the condition
$\lambda_{\text{FS}}\mathcal{F}_{0}^{1/2}=\mathcal{F}_{0}/4$. Thus,
$\lambda_{\text{FS}}=\mathcal{F}_{0}^{1/2}/4$ and Eq. (\ref{fsode}) becomes a
simple harmonic oscillator equation,%
\begin{equation}
\frac{d^{2}}{d\theta^{2}}q_{m}\left(  \theta\right)  +\frac{\mathcal{F}_{0}%
}{4}q_{m}\left(  \theta\right)  =0\text{.}\label{sho2}%
\end{equation}
The most general solution $q_{m}=q_{m}\left(  \theta\right)  $ of Eq.
(\ref{sho2}) is,%
\begin{equation}
q_{m}\left(  \theta\right)  =\cos\left(  v_{\mathcal{F}}\theta\right)
q_{m}\left(  0\right)  +\frac{\sin\left(  v_{\mathcal{F}}\theta\right)
}{v_{\mathcal{F}}}\dot{q}_{m}\left(  0\right)  \text{,}\label{GT2}%
\end{equation}
with $v_{\mathcal{F}}$ in Eq. (\ref{GT2}) defined as,%
\begin{equation}
v_{\mathcal{F}}\overset{\text{def}}{=}\mathcal{F}_{0}^{1/2}/2\text{.}%
\label{F2}%
\end{equation}
Finally, the Groverian probability path vector $\vec{p}\overset{\text{def}}%
{=}\vec{q}\cdot\vec{q}$ with $\vec{q}\overset{\text{def}}{=}(q_{1}$,...,
$q_{N})$ can be obtained by imposing the constraints $q_{\bar{w}}\left(
0\right)  =0$ and $\dot{q}_{\bar{w}}\left(  0\right)  =1$. These two
conditions yield $q_{\bar{w}}\left(  \theta\right)  =\sin\left(
\theta\right)  $ and $q_{j}\left(  \theta\right)  =\cos\left(  \theta\right)
$ for any $j\neq\bar{w}$ with $1\leq\bar{w}\leq N$. It\textbf{\ }therefore
follows that the success and failure probabilities for the quantum search
become $p_{\bar{w}}\left(  \theta\right)  =\sin^{2}\left(  \theta\right)  $
and $p_{w_{\perp}}\left(  \theta\right)  =\cos^{2}\left(  \theta\right)  $, respectively.

For the sake of clarity, we emphasize that Groverian geodesic paths are
referred to in this article as the continuous-time analogue of the output
quantum state emerging from Grover's algorithm viewed as the horizontal lift
of a geodesic in the complex projective Hilbert space. However, in the
geometric formulation of adiabatic quantum computation (AQC, \cite{farhi00}),
Groverian geodesic paths denote the time-optimal paths on a parameter manifold
for time-dependent control parameters that specify the time-dependent
interpolating Hamiltonian \textrm{H}$=\mathrm{H}\left[  \vec{x}\left(
t\right)  \right]  =\mathrm{H}\left[  1-x\left(  t\right)  \text{, }x\left(
t\right)  \right]  $ \cite{zanardi09}, where%
\begin{equation}
\mathrm{H}\left[  1-x\left(  t\right)  \text{, }x\left(  t\right)  \right]
\overset{\text{def}}{=}\left[  1-x\left(  t\right)  \right]  \mathrm{H}%
_{I}+x\left(  t\right)  \mathrm{H}_{P}\text{,}%
\end{equation}
with $0\leq t\leq T$. In the particular case of quantum search by local
adiabatic evolution \cite{roland02}, $\mathrm{H}_{I}=\mathrm{H}\left(
0\right)  \overset{\text{def}}{=}I-\left\vert s\right\rangle \left\langle
s\right\vert $ is the initial Hamiltonian, $\mathrm{H}_{P}=\mathrm{H}\left(
T\right)  \overset{\text{def}}{=}I-\left\vert w\right\rangle \left\langle
w\right\vert $ is the problem Hamiltonian, and $I$ denotes the identity
operator. In order to adiabatically drive the quantum system from the ground
state $\left\vert s\right\rangle $ of $\mathrm{H}_{I}$ (that is, the source
state) to a final state $\left\vert w\right\rangle $ that is close to the
ground state of $\mathrm{H}_{P}$ (that is, the target state) in the shortest
possible time, the optimal trajectory of the control parameter $x\left(
t\right)  $ is given by a Groverian geodesic path \cite{zanardi09,zanardi10},%
\begin{equation}
x\left(  \tau\right)  \overset{\text{def}}{=}\frac{1}{2}-\frac{1}{2\sqrt{N-1}%
}\tan\left[  \left(  1-2\tau\right)  \cos^{-1}\left(  \frac{1}{\sqrt{N}%
}\right)  \right]  \text{.} \label{aqc}%
\end{equation}
In Eq. (\ref{aqc}), $\tau$ denotes a dimensionless natural parameter
$\tau=\tau\left(  t\right)  $ with $\tau\left(  0\right)  =1$ and $\tau\left(
T\right)  =1$, while $N$ is the dimensionality of the search space.

\section{Harmonic geodesic motion beyond quantum settings}

In this Section, upon transitioning away the quantum setting, we explore the
motivations behind the emergence of SHO geodesic motions arising in general
relativistic and thermodynamical settings. In particular, we shall focus on
gravitational and thermal analogues of Eqs. (\ref{GT1}) and (\ref{GT2}).

\subsection{A gravitational example}

From a classical Newtonian mechanics perspective, a small mass that vibrates
under gravity about the center of a sphere composed of an ideal fluid will
exhibit a simple harmonic motion. For instance, consider an hydrometer of mass
$m$ that consists of a cylindrical stem of diameter $d$ and a spherical bulb
of volume $V_{0}$. Assume that the hydrometer is immersed in an ideal fluid of
constant density $\rho$ and the volume of displaced liquid is $V=V_{0}%
+\pi\left(  \frac{d}{2}\right)  ^{2}h$ with $h$ denoting the height of the
portion of the stem immersed in the liquid. At equilibrium, we have $mg=\rho
g\left[  V_{0}+\pi\left(  d/2\right)  ^{2}h\right]  $. When the hydrometer is
displaced by an additional distance $\Delta h$, the system is subject to a net
force $-\pi\rho g\left(  d/2\right)  ^{2}\Delta h$. Then, the system moves out
of equilibrium and begins floating/oscillating according to the dynamical
equation, $d^{2}\Delta h/dt^{2}+v_{\text{Newton}}^{2}\Delta h=0$, where
$v_{\text{Newton}}$ denotes the frequency of oscillation given by,%
\begin{equation}
v_{\text{Newton}}\overset{\text{def}}{=}\left(  \frac{\pi\left(  d/2\right)
^{2}}{m}\rho g\right)  ^{1/2}\text{.}%
\end{equation}
In what follows, motivated by the Zatzikis analysis in Ref. \cite{zatzkis59},
we show that such a point-like particle undergoes the same type of motion even
when the curvature of the space is considered, provided one assumes two
specific working conditions \cite{zatzkis59}: i) low velocities; ii) motion
limited to be confined near the center of the sphere. The equations of motion
of the small mass $m$ are,%
\begin{equation}
\frac{d^{2}x^{a}}{ds^{2}}+\Gamma_{bc}^{a}\frac{dx^{b}}{ds}\frac{dx^{c}}%
{ds}=0\text{,} \label{GRgeo}%
\end{equation}
with $\Gamma_{bc}^{a}$ denoting the Christoffel symbols of the second-kind and
$s$ in Eq. (\ref{GRgeo}) being an affine parameter \cite{defelice}. For
technical details on the possibility of using a non-affine parameter for
describing the geodesic motion, we refer to Appendix B. For the particular
physical scenario being analyzed \cite{zatzkis59}, we consider a stationary
physical system that exhibits spherical symmetry. In this case, the line
element $ds^{2}\overset{\text{def}}{=}g_{ab}\left(  x\right)  dx^{a}dx^{b}$
can be written as%
\begin{equation}
ds^{2}=g_{0}\left(  r\right)  dt^{2}-g_{1}\left(  r\right)  dr^{2}%
-r^{2}\left(  d\theta^{2}+\sin^{2}\theta d\varphi^{2}\right)  \text{,}
\label{statmech}%
\end{equation}
with $g_{0}\left(  r\right)  $ and $g_{1}\left(  r\right)  $ being\textbf{\ }%
time-independent quantities and $x^{a}\overset{\text{def}}{=}\left(  t\text{,
}r\text{, }\theta\text{, }\varphi\right)  $. The four geodesic relations in
Eq. (\ref{GRgeo}) reduce to two equations since we limit our analysis to
linear oscillations and, therefore, set $\theta=$\textrm{constant} and
$\varphi=$\textrm{constant}. We now focus on the interior of a sphere
composed\textbf{\ }of an ideal fluid (liquid) with constant density $\rho$.
Then, we assume that $g_{0}\left(  r\right)  $ and $g_{1}\left(  r\right)  $
are given by \cite{laue21},%
\begin{equation}
g_{0}\left(  r\right)  \overset{\text{def}}{=}\frac{1}{4}c^{2}\left[
3\cos\left(  \xi_{a}\right)  -\cos\left(  \xi\right)  \right]  \text{, and
}g_{1}\left(  r\right)  \overset{\text{def}}{=}\frac{1}{\cos^{2}\left(
\xi\right)  }\text{, } \label{supp1}%
\end{equation}
respectively, with $c$ being the speed of light. We remark that the quantities
$\xi$ and $\xi_{a}$ in\ Eq. (\ref{supp1}) are defined as%
\begin{equation}
\xi\overset{\text{def}}{=}\sin^{-1}\left(  \frac{r}{R}\right)  \text{, and
}\xi_{a}\overset{\text{def}}{=}\sin^{-1}\left(  \frac{a}{R}\right)  \text{,}
\label{supp2}%
\end{equation}
respectively, with $R\overset{\text{def}}{=}c\left[  3/\left(  8\pi
G\rho\right)  \right]  ^{1/2}$ where $\rho$ is the constant density of the
ideal (that is, incompressible) fluid, $G$ is Newton's universal gravitational
constant, and $a$ is the value of $r$ on the surface of the sphere (thus, it
represents the maximal value of $r$). By substituting Eqs. (\ref{supp1}) and
(\ref{supp2}) into Eq. (\ref{GRgeo}), recalling that we assumed $\theta
=$\textrm{constant} and $\varphi=$\textrm{constant} and finally, taking the
limit of low velocities ($\dot{\xi}^{2}\ll1$) together with considering
dynamical changes near the center of the sphere ($\sin\left(  \theta\right)
\approx\theta$ and $\cos\left(  \theta_{a}\right)  \approx1$), the classical
equations of motion for $\xi\left(  t\right)  $ becomes the familiar simple
harmonic oscillator equation%
\begin{equation}
\frac{d^{2}\xi}{dt^{2}}+v_{\text{GR}}^{2}\xi=0\text{.} \label{grsho}%
\end{equation}
The most general solution of Eq. (\ref{grsho}) can be written as,%
\begin{equation}
\xi\left(  t\right)  =\cos\left(  v_{\text{GR}}t\right)  \xi\left(  0\right)
+\frac{\sin\left(  v_{\text{GR}}t\right)  }{v_{\text{GR}}}\dot{\xi}\left(
0\right)  \text{,} \label{GT3}%
\end{equation}
where the angular frequency $v_{\text{GR}}$ in Eq. (\ref{grsho}) is a constant
quantity given by,%
\begin{equation}
v_{\text{GR}}\overset{\text{def}}{=}\left(  \frac{4\pi}{3}G\rho\right)
^{1/2}\text{.} \label{vgr}%
\end{equation}
As a final remark, we point out that the constancy of $v_{\text{GR}}$ in Eq.
(\ref{vgr}) is a consequence of the constancy of $\rho$. The latter property
arises due to our consideration of an ideal fluid. Finally, the\ working
assumption of \textquotedblleft ideality\textquotedblright\ enabled us to
consider a stationary metric in Eq. (\ref{statmech}).

\subsection{A thermal example}

In what follows, we consider a famous experiment in thermodynamics where a
simple harmonic oscillatory motion occurs. Specifically, we take into account
the R\"{u}chhardt experiment used to measure the adiabatic coefficient
$\gamma$ of an ideal gas \cite{ruchhardt29}. Recall that $\gamma
\overset{\text{def}}{=}C_{P}/C_{V}$ is the ratio between the heat capacity at
constant pressure ($C_{P}$) and the heat capacity at constant volume ($C_{V}%
$). The experiment is described by an adiabatic ($PV^{\gamma}=$%
\textrm{constant}) compression of a mole of an ideal gas ($PV=RT$). The
quantities $P$, $V$, and $T$ denote pressure, volume, and temperature,
respectively. The quantity $R$\textbf{\ }is the universal gas constant. We
recall that the relation $PV^{\gamma}=$\textrm{constant }is a consequence of
the first principle of thermodynamics ($dU=-PdV+dQ$) applied to one mole of an
ideal gas ($dU=C_{V}dT$) in the absence of any heat exchange with the external
environment ($dQ=0$) \cite{fermi94}. The quantities $U$ and $Q$ denote energy
and heat, respectively.

We remark at this juncture that the R\"{u}chhardt experiment can be explained
via the following sequence of steps \cite{ruchhardt29}. A spherical piston of
mass $m$ is allowed to fall under uniform gravity in a cylindrical tube of
volume $V$ and of cross-section $A$ which is open on one of its end. The
piston has a uniform cross-section so as to create an air-tight seal. The gas
trapped within the cylindrical tube is adiabatically compressed by the weight
of the spherical piston. During this compression, the temperature of the gas
increases. Furthermore\textbf{,} as the piston falls, the piston experiences
bounce due to the creation of a gas cushion. As a consequence, the piston
begins exhibiting simple harmonic oscillations. At equilibrium, the pressure
$P_{0}$ of the trapped gas within the tube equals the sum of the atmospheric
pressure ($P_{\mathrm{atm}}$) and the pressure ($mg/A$) exerted by the piston
on the gas. As the piston moves away from equilibrium by an infinitesimal
distance $x$, the pressure changes by $dP$. Then, the Newtonian equation of
motion projected along the $x$-axis, $F_{x}=m\ddot{x}=AdP$, for the piston
becomes%
\begin{equation}
\frac{d^{2}x}{dt^{2}}+v_{\text{th}}^{2}x=0\text{,} \label{tsho}%
\end{equation}
where $x$ denotes the position of the piston. The most general solution of Eq.
(\ref{tsho}) can be written as,%
\begin{equation}
x\left(  t\right)  =\cos\left(  v_{\text{th}}t\right)  x\left(  0\right)
+\frac{\sin\left(  v_{\text{th}}t\right)  }{v_{\text{th}}}\dot{x}\left(
0\right)  \text{,} \label{GT4}%
\end{equation}
where the angular frequency of oscillations $v_{\text{th}}$ in Eq.
(\ref{tsho}) is a constant quantity given by,%
\begin{equation}
v_{\text{th}}\overset{\text{def}}{=}\left(  \frac{P_{0}A^{2}}{mV_{0}}%
\gamma\right)  ^{1/2}\text{,} \label{omega}%
\end{equation}
with $P_{0}$ and $V_{0}$ being the equilibrium pressure and volume,
respectively. The adiabatic coefficient $\gamma$ can be experimentally
determined by measuring the period of oscillations of the piston,
$T\overset{\text{def}}{=}2\pi/v_{\text{th}}$. We emphasize that $v_{\text{th}%
}$ in Eq. (\ref{omega}) is constant since $\gamma$ is assumed to be constant.
In general, for all ideal gases, $\gamma>1$ and depends on the temperature
\cite{zemansky97}. Moreover, for monotonic gases (for instance, \textrm{He}
and \textrm{Ne}), $\gamma$ is constant over a wide range of temperatures. For
diatomic (for instance, \textrm{H}$_{2}$ and \textrm{O}$_{2}$) and polyatomic
gases (for instance, \textrm{CO}$_{2}$ and \textrm{NH}$_{3}$) however,
$\gamma$ varies with $T$. In particular, a very large change in temperature
can produce a non negligible change in $\gamma$. However, for (quasi-static)
adiabatic thermodynamic processes characterized by a small temperature change,
the change in $\gamma$ can be neglected and $\gamma$ can be regarded as a
constant quantity. In the (quasi-static) adiabatic compression considered
here, although the temperature of the gas increases, we assume that the
overall change in temperature is so small that $\gamma$ can be essentially
considered to be\textbf{\ }a constant quantity over this temperature range.

As a final observation, we point out that the method of measuring $\gamma$
developed by R\"{u}chhardt requires only the use of classical Newtonian
mechanics. This in turn, can also be presented in a Riemannian geometric
fashion once we consider the Hamiltonian dynamical formulation of Newton's
construction \cite{casetti00}. Indeed, consider a conservative Hamiltonian
system specified by an Hamiltonian $\mathrm{H}\left(  p\text{, }q\right)
\overset{\text{def}}{=}p^{2}/(2m)+V\left(  q\right)  $ with\textbf{ }%
$p$\textbf{ }and $q$\textbf{ }being the generalized momentum and coordinate,
respectively, while the energy\textbf{ }$E$\textbf{ }is a conserved quantity.
Then, its dynamics can be recast in terms of geodesic motion on a Riemannian
manifold. Such a manifold is specified by the configuration space of the
dynamical system being considered and is equipped with a metric structure
defined by a (conformally flat) Jacobi metric tensor $g_{ab}$ defined as
\cite{casetti00},%
\begin{equation}
g_{ab}\left(  q\right)  \overset{\text{def}}{=}2\left[  E-V\left(  q\right)
\right]  \delta_{ab}\text{,} \label{Jacobi}%
\end{equation}
where $1\leq a$, $b\leq n_{\text{\textrm{dof}}}$ with $n_{\text{\textrm{dof}}%
}$ being equal to the number of degrees of freedom of the dynamical system.
The geodesic equations,%
\begin{equation}
\frac{d^{2}q^{a}}{ds^{2}}+\Gamma_{bc}^{a}\frac{dq^{b}}{ds}\frac{dq^{c}}%
{ds}=0\text{,} \label{geoshit}%
\end{equation}
can be obtained by minimizing the action functional (that is, the length)
$\mathrm{S}\left[  q\right]  $,%
\begin{equation}
\mathrm{S}\left[  q\right]  \overset{\text{def}}{=}\int\left[  g_{ab}\left(
q\right)  dq^{a}dq^{b}\right]  ^{1/2}\text{,}%
\end{equation}
with $q\overset{\text{def}}{=}\left(  q^{1}\text{,..., }%
q^{n_{\text{\textrm{dof}}}}\right)  $ being local coordinates on the curved
manifold. For the sake of clarity, we emphasize that $s$ in Eq. (\ref{geoshit}%
) is the arc-length parameter and is related to the physical time $t$ via the
relation $ds=2Tdt$, with $T$ being the kinetic energy of the physical system.
Finally, the Riemannian geometrization of R\"{u}chhardt classical mechanical
description of the measurement yielding the adiabatic coefficient of an ideal
gas can be obtained in a straightforward manner once we observe that
$n_{\text{\textrm{dof}}}=1$, $q=x$, and the potential $V\left(  q\right)  $ in
Eq. (\ref{Jacobi}) reduces to the harmonic potential $(1/2)mv_{\text{th}}%
^{2}x^{2}$ with $v_{\text{th}}$ defined in Eq. (\ref{omega}). We refer to
Appendix C for an explicit derivation of Newton's equation of motion in Eq.
(\ref{tsho}) starting from Eq. (\ref{geoshit}). Finally, for further
discussions on the role played by conformally flat Jacobi metrics in physics,
we refer to Refs. \cite{caticha07,cafaro09}.

We have discussed in this section the emergence of simple harmonic motion in
gravitational and thermodynamical settings. In general, the link between
gravity and thermodynamics is a rather fascinating topic \cite{padmanabhan}.
For a specific perspective on a cosmological model of dark energy using an
adiabatic fluid satisfying the relation $PV^{\gamma}=\mathrm{constant}$ with a
constant adiabatic coefficient $\gamma$ and evolving according to classical
laws of thermodynamics, we refer to Ref. \cite{luongo16}.

\begin{table}[t]
\centering
\begin{tabular}
[c]{c|c|c|c|c}\hline\hline
Type of Physics & Type of Metric & Geodesic Equation & Frequency of SHO &
Conserved Quantity\\\hline
quantum & Fubini-Study & state vector & $\Delta E/\hslash$ & $\Delta E$,
energy dispersion\\
quantum information & Fubini-Study & probability amplitude & $\mathcal{F}%
^{1/2}/2$ & $\mathcal{F}$, Fisher information\\
gravitational & curved Lorentzian spacetime & radial position of particle &
$\left(  \frac{4\pi G}{3}\rho\right)  ^{1/2}$ & $\rho$, ideal liquid mass
density\\
thermal & Jacobi & position of piston & $\left(  \frac{P_{0}A^{2}}{mV_{0}%
}\gamma\right)  ^{1/2}$ & $\gamma$, ideal gas adiabatic coefficient\\\hline
\end{tabular}
\caption{Schematic representation of the emergence of SHO motion from
different physics frameworks, diverse metric structures, distinct variables
satisfying the geodesic equation, various frequencies and, finally,
distinctive conserved quantities. The common theme in each and every physical
scenario is the presence of a peculiar conserved quantity.}%
\end{table}

\section{Concluding remarks}

In this article, we presented a unifying variational calculus derivation of
Groverian geodesics for both {the horizontal lift of} quantum state vectors
(see Eqs. (\ref{GT1}) and (\ref{F1})) as well as quantum probability
amplitudes (see Eqs. (\ref{GT2}) and (\ref{F2})). In the first case, following
the Mukunda-Simon work in Ref. \cite{mukunda93}, we demonstrated that
horizontal affinely parametrized {lift} of geodesic paths on the manifold of
Hilbert space rays arise from the minimization of the length specified by the
Fubini-Study metric (see Eq. (\ref{FSyes})). In the second case, inspired by
the Alvarez-Gomez work in Ref. \cite{alvarez00}, we explicitly illustrated
that geodesic paths for probability amplitudes emerge as a consequence of
minimizing the length expressed in terms of the Fisher information (see Eq.
(\ref{fessa}) with $ds_{\text{FS}}^{2}\overset{\text{def}}{=}(1/4)\mathcal{F}%
\left(  \theta\right)  d\theta^{2}$). In both derivations, we note that
geodesic equations are described by simple harmonic oscillators (SHOs). While
in the first derivation, the frequency of oscillations $v_{\mathrm{H}}$ is
proportional to the (constant) energy dispersion $\Delta E$ of the Hamiltonian
system; in the second derivation the frequency of oscillations $v_{\mathcal{F}%
}$ is proportional to the square-root $\sqrt{\mathcal{F}}$ of the (constant)
Fisher information. Interestingly, by equating the two frequencies in Eqs.
(\ref{F1}) and (\ref{F2}), we are able to recover the well-known
Anandan-Aharonov relation connecting the squared speed $v_{\mathrm{H}}^{2}$ of
evolution of an Hamiltonian system with its energy dispersion $\Delta E$,
together with the concise link between speed of evolution $v_{\mathrm{H}}$ and
Fisher information $\mathcal{F}$,%
\begin{equation}
\frac{\Delta E^{2}}{\hslash^{2}}=v_{\mathrm{H}}^{2}=\frac{\mathcal{F}}%
{4}\text{,}%
\end{equation}

We point out that the emergence of simple harmonic motion in our geometrical
investigations of quantum mechanical phenomena is not entirely unexpected. For
instance, the normalization conditions linked to the probabilistic nature of
quantum mechanics, namely $\left\langle \psi|\psi\right\rangle =1$ and $q\cdot
q=1$, have played an important role in the derivations of geodesic
trajectories for state vectors and probability amplitudes, respectively.
Stated otherwise, in both cases the trajectories were constrained to be on a
spherical surface. In the framework of classical Newtonian mechanics, simple
harmonic motion can be shown to emerge from the study of a (free)
point-particle of mass $m$ subject to no external force but constrained to
move an a spherical surface of constant radius $R_{0}$. The dynamical
trajectory $x\left(  t\right)  $ of such a particle on a sphere defined by the
condition $x\cdot x=R_{0}^{2}$ can be obtained from the Euler-Lagrange
equations emerging from the constrained Lagrangian $\mathcal{L}\left(  \dot
{x}\text{, }x\text{, }t\right)  \overset{\text{def}}{=}\left(  m/2\right)
\left(  \dot{x}\cdot\dot{x}\right)  -\Lambda\left(  t\right)  \left[
1-\left(  x\cdot x\right)  /R_{0}^{2}\right]  $ where $x=x\left(  t\right)  $,
$\dot{x}\overset{\text{def}}{=}dx/dt$, with $\Lambda\left(  t\right)  $ being
a Lagrange multiplier. We observe that although Lagrange multipliers are
generally time-dependent quantities \cite{goldstein}, we could have scenarios
where one deals with space-time-dependent Lagrange multipliers. This happens,
for instance, in time-dependent information constrained optimization problems
such as those tackled with MaxCal inference algorithms
\cite{jaynes85,cafaro16}. Having said that, the trajectory $x\left(  t\right)
$ is described by a great circle, that is, a geodesic on the sphere along
which the free particle moves with constant speed (that is, the magnitude of
the velocity vector). Interestingly, in such a classical mechanical scenario,
the Lagrange multiplier $\Lambda\left(  t\right)  $ is determined to be equal
to the time-independent (that is, constant) kinetic energy of the particle,
$T\overset{\text{def}}{=}(1/2)mv^{2}$. We also emphasize that simple harmonic
motion does not arise merely from the condition of constrained motion on a
sphere. Instead, it can be shown that the constancy of some relevant physical
quantity must to be satisfied. In particular, in the three examples being
compared in this discussion, the constancy of the kinetic energy $T$ (that is,
$v_{\text{geo-Newton}}\propto\sqrt{T}$) in the classical scenario replaces the
constancy of the energy dispersion $\Delta E$ (that is, $v_{\text{geo-QM}%
}\propto\sqrt{\Delta E^{2}}$); in the investigation of geodesics in quantum
ray spaces and that of the Fisher information $\mathcal{F}$ (that is,
$v_{\text{geo-Grover}}\propto\sqrt{\mathcal{F}}$) in the geometric analysis of
Grover's algorithm. As a concluding remark in this specific set of remarks, we
point out that further enlightening discussions concerning geodesic motion on
spheres of relevance in quantum mechanics can be found in Ref. \cite{karol}.

Finally, upon transitioning away from the quantum setting, we discuss the
universality of the emergence of geodesic motion of SHO type in the presence
of conserved quantities by analyzing two specific phenomena of gravitational
(see Eqs. (\ref{GT3}) and (\ref{vgr})) and thermodynamical origin (see Eqs.
(\ref{GT4}) and (\ref{omega})), respectively. In the gravitational case, the
harmonic motion emerges from consideration of the geodesic motion on a curved
manifold with Lorentzian spacetime metric (see Eq.(\ref{statmech})). In the
thermodynamical case, by contrast, the harmonic motion emerges by considering
the geodesic motion on a configuration manifold equipped with a conformally
flat Jacobian metric (see Eq.(\ref{Jacobi})). A global summary of the various
features concerning the four physical scenarios investigated in the present
article appears in Table I.

Despite the pedagogical nature of our variational calculus reconsideration of
Groverian paths for both state vectors and probability amplitudes, the
decision to discuss them in parallel enabled us to clarify the physical
connection among energy dispersion ($\Delta E$), speed of evolution
($v_{\mathrm{H}}$) of the system, and the Fisher information ($\mathcal{F}$).
For the sake of transparency, we emphasize that the physical relevance of the
pairs ($\Delta E$, $v_{\mathrm{H}}$), ($\mathcal{F}$, $v_{\mathrm{H}}$), and
($\Delta E$, $\mathcal{F}$) has been previously described in contexts
different from ours in Refs. \cite{anandan90}, \cite{brody96,pezze09,taddei13}%
, and \cite{boixo07}, respectively. Our work is however, unique in the sense
that we provide a unifying physical link for $\left(  \Delta E\text{,
}v_{\mathrm{H}}\text{, }\mathcal{F}\right)  $ in the novel context of a
geometric characterization of quantum searching with the underlying physical
motivation of finding a good geometric measure of thermodynamic efficiency for
very fast quantum transfer phenomena \cite{carlopre,carlopre20}. Moreover, by
highlighting the \textquotedblleft universality\textquotedblright\ of harmonic
geodesic motion in the presence of conserved quantities associated with
physical contexts other than the quantum one, we believe our work can help
characterize realistic deviations from ideal Groverian paths in quantum
searching by mimicking departures from the harmonic condition in more
realistic physical settings \cite{carlopre,carlopre20}.

While our current degree of rational belief requires further physical and
mathematical justification, these types of theoretical analogies could well
serve as the nascent forms that ultimately lead to significant findings such
as the famous link between optimization methods and annealing in solids
\cite{vecchi83}. We remain, as ever, highly motivated to further develop these
avenues of investigation in future scientific efforts.

\begin{acknowledgments}
C.C. is grateful to the United States Air Force Research Laboratory (AFRL)
Summer Faculty Fellowship Program for providing support for this work. P.M.A.
acknowledges support from the Air Force Office of Scientific Research (AFOSR).
Any opinions, findings and conclusions or recommendations expressed in this
material are those of the author(s) and do not necessarily reflect the views
of the Air Force Research Laboratory (AFRL).
\end{acknowledgments}

\bigskip\pagebreak

\appendix

\section{Geodesics in the space of unit rays}

In this Appendix, we present technical details on the derivation of horizontal
affinely parametrized geodesics \cite{mukunda93}. This derivation was omitted
in Section III. We also briefly discuss the notion of non-horizontal,
non-affinely parametrized geodesics in the second part of this Appendix.

\subsection{Horizontal affinely parametrized geodesics}

{Let $\mathcal{H}$ be an $N$-dimensional Hilbert space endowed with the scalar
product $\langle\cdot\,|\,\cdot\rangle$. Consider the subset $\mathcal{H}%
_{0}\subset\mathcal{H}$ of unitary vectors in $\mathcal{H}$, i.e. $\psi
\in\mathcal{H}_{0}$ if $\langle\psi\,|\,\psi\rangle=1$. A one-parameter smooth
curve $\widetilde{\gamma}_{0}$ in $\mathcal{H}_{0}$ consists of a family of
vectors $\psi(s)$:
\begin{equation}
\widetilde{\gamma}_{0}\overset{\text{def}}{=}\left\{  \psi(s)\in
\mathcal{H}_{0}\,|\,s\in\lbrack s_{1},s_{2}]\in\mathbb{R}\right\}
\,.\label{curveH0}%
\end{equation}
From the assumption that $\langle\psi(s)\,|\,\psi(s)\rangle=1$, }it
immediately follows{\textbf{ }that $\mbox{Re}\langle\psi(s)\,|\,\dot{\psi
}(s)\rangle=0$, }with{ $\dot{\psi}(s)=\frac{\partial}{\partial\,s}\psi(s)$.
This }result is{ equivalent to
\begin{equation}
\langle\psi(s)\,|\,\dot{\psi}(s)\rangle=i\mbox{Im}\langle\psi(s)\,|\,\dot
{\psi}(s)\rangle\,.\label{ImaginaryH0}%
\end{equation}
A }gauge transformation{ of $\widetilde{\gamma}_{0}$ is typically determined
by a \emph{real }}phase factor{ $\alpha(s)$ by taking the curve $\widetilde
{\gamma}_{0}$ into a new one $\widetilde{\gamma}_{0}^{\prime}$:
\begin{equation}
\widetilde{\gamma}_{0}\rightarrow\widetilde{\gamma}_{0}^{\prime}\,,\quad
\psi^{\prime}(s)=e^{i\,\alpha(s)}\,\psi(s)\,,\quad s\in\lbrack s_{1}%
,s_{2}]\,.\label{gaugetransf}%
\end{equation}
This transformation applies to Eq. \eqref{ImaginaryH0} as follows:
\begin{equation}
\mbox{Im}\langle\psi^{\prime}(s)\,|\,\dot{\psi}^{\prime}(s)\rangle
=\mbox{Im}\langle\psi(s)\,|\,\dot{\psi}(s)\rangle+\dot{\alpha}%
(s)\,.\label{ImGauge}%
\end{equation}
This allows us to construct a functional of $\widetilde{\gamma}_{0}$ which is
gauge invariant:
\begin{equation}
\mbox{arg}(\psi^{\prime}(s_{1})\,|\,\psi^{\prime}(s_{2}))-\mbox{Im}\int
_{s_{1}}^{s_{2}}\,\langle\psi^{\prime}(s)\,|\,\dot{\psi}^{\prime}%
(s)\rangle\,ds=\mbox{arg}(\psi(s_{1})\,|\,\psi(s_{2}))-\mbox{Im}\int_{s_{1}%
}^{s_{2}}\,\langle\psi(s)\,|\,\dot{\psi}(s)\rangle\,ds\,,\label{Gaugeinv}%
\end{equation}
where $\mbox{arg}(\psi^{\prime}(s_{1})\,|\,\psi^{\prime}(s_{2}))=\int_{s_{1}%
}^{s_{2}}\dot{\alpha}(s)\,ds$. This property can be interpreted in\textbf{ }%
}the space of unit rays{ $\mathcal{H}\mathbf{P}^{N}$}$\simeq%
\mathbb{C}
\mathcal{P}^{N-1}${, which is }simply{ the quotient of $\mathcal{H}_{0}$ under
the action $\psi^{\prime}=e^{i\alpha}\psi$ with $\alpha\in\mathbb{R}$}. {If we
denote by $\pi$ the corresponding projection map:
\begin{equation}
\pi:\mathcal{H}_{0}\rightarrow\mathcal{H}\mathbf{P}^{N}\,,\quad\pi
(\widetilde{\gamma})=\pi(e^{i\alpha}\widetilde{\gamma})\,,\quad
\mbox{for all}\,\,\alpha\in\mathbb{R}\,\,\,\mbox{and}\,\,\,\psi\in
\mathcal{H}_{0}\,,\label{projection}%
\end{equation}
we }observe{ that the curve $\widetilde{\gamma}_{0}$ projects onto a smooth
curve $\gamma_{0}$ in $\mathcal{H}\mathbf{P}^{N}$. Moreover, if $\widetilde
{\gamma}_{0}^{\prime}$ is obtained from $\widetilde{\gamma}_{0}$ by the
transformation \eqref{gaugetransf} we }find{ that $\pi(\widetilde{\gamma}%
_{0}^{\prime})=\gamma_{0}$. }

The gauge invariance of \eqref{Gaugeinv} implies that it is actually a
functional of $\gamma_{0}$. Another important property of the functional
\eqref{Gaugeinv} is that it is reparametrization invariant. Combining these
two properties, we can write
\begin{equation}
\varphi\lbrack\gamma_{0}]=\mbox{arg}\langle\psi(s_{1})\,|\,\psi(s_{2}%
)\rangle-\mbox{Im}\int_{s_{1}}^{s_{2}}\langle\psi(s^{\prime})\,|\,\dot{\psi
}(s^{\prime})\rangle\,ds^{\prime}\,, \label{raysfunctional}%
\end{equation}
which is called the \emph{geometric phase} associated with the smooth curve
$\gamma_{0}$. Now, given $\gamma_{0}\in\mathcal{H}\mathbf{P}^{N}$, the
\textit{horizontal lift} $\widetilde{\gamma}_{0}$ of $\gamma_{0}$ is specified
by requiring that
\begin{equation}
\mbox{Im}\langle\psi(s^{\prime})\,|\,\dot{\psi}(s^{\prime})\rangle
=0\quad\Leftrightarrow\quad\langle\psi(s^{\prime})\,|\,\dot{\psi}(s^{\prime
})\rangle=0\,.
\end{equation}
We can see from Eq. \eqref{raysfunctional} that this requirement implies that
for a given $\gamma_{0}$ and a given initial $\psi(s_{1})$ projecting onto the
initial point of $\gamma_{0}$ there is a unique horizontal lift of $\gamma
_{0}$ starting from $\psi(s_{1})$. In particular, this allows us to follow
geodesic paths in $\mathcal{H}\mathbf{P}^{N}$ by studying the behavior of
their horizontal lift in the Hilbert space of unit vectors.

We recall that geodesic paths $\gamma_{\mathrm{geo}}\left(  s\right)  $ in
$\mathcal{H}\mathbf{P}^{N}$, with $s_{1}\leq s\leq s_{2}$, are those for which
the action functional $\mathcal{L}\left[  \gamma_{\mathrm{geo}}\right]  $
(that is, the length),%
\begin{equation}
\mathcal{L}\left[  \gamma_{\mathrm{geo}}\right]  \overset{\text{def}}{=}%
\int_{s_{1}}^{s_{2}}\sqrt{dl_{\text{FS}}^{2}}=\int_{s_{1}}^{s_{2}}\left\langle
u_{\perp}\left(  s\right)  |u_{\perp}\left(  s\right)  \right\rangle
^{1/2}ds\text{,} \label{action1}%
\end{equation}
is stationary. Recalling that $\left\vert u_{\perp}\right\rangle
\overset{\text{def}}{=}\left\vert u\right\rangle -\left\langle \psi
|u\right\rangle \left\vert \psi\right\rangle $ with $\left\vert u\right\rangle
\overset{\text{def}}{=}\left\vert \dot{\psi}\right\rangle $ and $\left\langle
\psi|\psi\right\rangle =1$, $\mathcal{L}\left[  \gamma_{\mathrm{geo}}\right]
$ in Eq. (\ref{action1}) can be rewritten as,%
\begin{equation}
\mathcal{L}\left[  \gamma_{\mathrm{geo}}\right]  =\int_{s_{1}}^{s_{2}}\left[
\left\langle u|u\right\rangle -\left\langle u|\psi\right\rangle \left\langle
\psi|u\right\rangle \right]  ^{1/2}ds\text{.} \label{action2}%
\end{equation}
Recalling further that $z+z^{\ast}=2\operatorname{Re}\left(  z\right)  $ for
any $z\in%
\mathbb{C}
$, the variation $\delta\mathcal{L}\left[  \gamma_{\mathrm{geo}}\right]  $ of
$\mathcal{L}\left[  \gamma_{\mathrm{geo}}\right]  $ in Eq. (\ref{action2})
becomes,
\begin{equation}
\delta\mathcal{L}\left[  \gamma_{\mathrm{geo}}\right]  =\int_{s_{1}}^{s_{2}%
}\frac{1}{\left\Vert \left\vert u_{\perp}\right\rangle \right\Vert
}\operatorname{Re}\left[  \left\langle \delta u|u_{\perp}\right\rangle
-\left\langle \delta\psi|u\right\rangle \left\langle u|\psi\right\rangle
\right]  ds\text{.} \label{action3}%
\end{equation}
At this juncture, we emphasize that the quantity $-\operatorname{Re}\left[
\left\langle \delta\psi|u\right\rangle \left\langle u|\psi\right\rangle
\right]  $ in Eq. (\ref{action3}) equals $\operatorname{Re}\left[
\left\langle \delta\psi|u_{\perp}\right\rangle \left\langle \psi
|u\right\rangle \right]  $. This equality is a consequence of the fact that
the inner products $\left\langle \delta\psi|\psi\right\rangle $ and
$\left\langle u|\psi\right\rangle $ are pure imaginary numbers. These
relations, in turn, are ultimately a consequence of the normalization
condition $\left\langle \psi|\psi\right\rangle =1$. More explicitly, observe
that%
\begin{align}
\operatorname{Re}\left[  \left\langle \delta\psi|u_{\perp}\right\rangle
\left\langle \psi|u\right\rangle +\left\langle \delta\psi|u\right\rangle
\left\langle u|\psi\right\rangle \right]   &  =\operatorname{Re}\left[
\left\langle \delta\psi|u_{\perp}\right\rangle \left\langle u|\psi
\right\rangle ^{\ast}+\left\langle \delta\psi|u\right\rangle \left\langle
u|\psi\right\rangle \right] \nonumber\\
&  =\operatorname{Re}\left[  -\left\langle \delta\psi|u_{\perp}\right\rangle
\left\langle u|\psi\right\rangle +\left\langle \delta\psi|u\right\rangle
\left\langle u|\psi\right\rangle \right] \nonumber\\
&  =\operatorname{Re}\left\{  \left\langle u|\psi\right\rangle \left[
-\left\langle \delta\psi|u_{\perp}\right\rangle +\left\langle \delta
\psi|u\right\rangle \right]  \right\} \nonumber\\
&  =\operatorname{Re}\left[  -\left\langle u|\psi\right\rangle \left\langle
\delta\psi|\psi\right\rangle \left\langle \psi|u\right\rangle \right]
\nonumber\\
&  =\operatorname{Re}\left[  -\left\vert \left\langle u|\psi\right\rangle
\right\vert ^{2}\left\langle \delta\psi|\psi\right\rangle \right]  \text{.}
\label{wrong}%
\end{align}
Thus, since $\left\vert \left\langle u|\psi\right\rangle \right\vert ^{2}$ is
real and $\left\langle \delta\psi|\psi\right\rangle $ is purely imaginary, we
obtain from Eq. (\ref{wrong}) the result\textbf{\ }that $\operatorname{Re}%
\left[  \left\langle \delta\psi|u_{\perp}\right\rangle \left\langle
\psi|u\right\rangle \right]  =-\operatorname{Re}\left[  \left\langle
\delta\psi|u\right\rangle \left\langle u|\psi\right\rangle \right]  $.
Therefore, by\textbf{\ }exploiting this last relation, $\delta\mathcal{L}%
\left[  \gamma_{\mathrm{geo}}\right]  $ in Eq. (\ref{action3}) becomes%
\begin{equation}
\delta\mathcal{L}\left[  \gamma_{\mathrm{geo}}\right]  =\int_{s_{1}}^{s_{2}%
}\frac{1}{\left\Vert \left\vert u_{\perp}\right\rangle \right\Vert
}\operatorname{Re}\left[  \left\langle \delta u|u_{\perp}\right\rangle
+\left\langle \delta\psi|u_{\perp}\right\rangle \left\langle \psi
|u\right\rangle \right]  ds\text{,}%
\end{equation}
that is,%
\begin{equation}
\delta\mathcal{L}\left[  \gamma_{\mathrm{geo}}\right]  =\int_{s_{1}}^{s_{2}%
}\operatorname{Re}\left[  \left\langle \frac{d}{ds}\left(  \delta\psi\right)
|\frac{\left\vert u_{\perp}\right\rangle }{\left\Vert \left\vert u_{\perp
}\right\rangle \right\Vert }\right\rangle \right]  ds+\int_{s_{1}}^{s_{2}%
}\operatorname{Re}\left[  \left\langle \delta\psi|\frac{\left\vert u_{\perp
}\right\rangle }{\left\Vert \left\vert u_{\perp}\right\rangle \right\Vert
}\right\rangle \left\langle \psi|u\right\rangle \right]  ds\text{.}
\label{spandau}%
\end{equation}
Observe that by eliminating the boundary terms, the first term on the RHS of
Eq. (\ref{spandau}) can be written as%
\begin{align}
\int_{s_{1}}^{s_{2}}\operatorname{Re}\left[  \left\langle \frac{d}{ds}\left(
\delta\psi\right)  |\frac{\left\vert u_{\perp}\right\rangle }{\left\Vert
\left\vert u_{\perp}\right\rangle \right\Vert }\right\rangle \right]  ds  &
=\int_{s_{1}}^{s_{2}}\operatorname{Re}\left\{  \frac{d}{ds}\left[
\left\langle \delta\psi|\frac{\left\vert u_{\perp}\right\rangle }{\left\Vert
\left\vert u_{\perp}\right\rangle \right\Vert }\right\rangle \right]
\right\}  ds-\int_{s_{1}}^{s_{2}}\operatorname{Re}\left[  \left\langle
\delta\psi|\frac{d}{ds}\frac{\left\vert u_{\perp}\right\rangle }{\left\Vert
\left\vert u_{\perp}\right\rangle \right\Vert }\right\rangle \right]
ds\nonumber\\
&  =\left\{  \operatorname{Re}\left[  \left\langle \delta\psi|\frac{\left\vert
u_{\perp}\right\rangle }{\left\Vert \left\vert u_{\perp}\right\rangle
\right\Vert }\right\rangle \right]  \right\}  _{s_{1}}^{s_{2}}-\int_{s_{1}%
}^{s_{2}}\operatorname{Re}\left[  \left\langle \delta\psi|\frac{d}{ds}%
\frac{\left\vert u_{\perp}\right\rangle }{\left\Vert \left\vert u_{\perp
}\right\rangle \right\Vert }\right\rangle \right]  ds\nonumber\\
&  =-\int_{s_{1}}^{s_{2}}\operatorname{Re}\left[  \left\langle \delta
\psi|\frac{d}{ds}\frac{\left\vert u_{\perp}\right\rangle }{\left\Vert
\left\vert u_{\perp}\right\rangle \right\Vert }\right\rangle \right]
ds\text{,}%
\end{align}
that is,%
\begin{equation}
\int_{s_{1}}^{s_{2}}\operatorname{Re}\left[  \left\langle \frac{d}{ds}\left(
\delta\psi\right)  |\frac{\left\vert u_{\perp}\right\rangle }{\left\Vert
\left\vert u_{\perp}\right\rangle \right\Vert }\right\rangle \right]
ds=-\int_{s_{1}}^{s_{2}}\operatorname{Re}\left[  \left\langle \delta\psi
|\frac{d}{ds}\frac{\left\vert u_{\perp}\right\rangle }{\left\Vert \left\vert
u_{\perp}\right\rangle \right\Vert }\right\rangle \right]  ds\text{.}
\label{jlo}%
\end{equation}
By combining Eqs. (\ref{spandau}) and (\ref{jlo}), the variation
$\delta\mathcal{L}\left[  \gamma_{\mathrm{geo}}\right]  $ becomes,%
\begin{equation}
\delta\mathcal{L}\left[  \gamma_{\mathrm{geo}}\right]  =-\int_{s_{1}}^{s_{2}%
}\operatorname{Re}\left[  \left\langle \delta\psi|\frac{d}{ds}\frac{\left\vert
u_{\perp}\right\rangle }{\left\Vert \left\vert u_{\perp}\right\rangle
\right\Vert }-\left\langle \psi|u\right\rangle \frac{\left\vert u_{\perp
}\right\rangle }{\left\Vert \left\vert u_{\perp}\right\rangle \right\Vert
}\right\rangle \right]  ds\text{.} \label{action4}%
\end{equation}
Upon imposing the requirement that $\delta\mathcal{L}\left[  \gamma
_{\mathrm{geo}}\right]  =0$ for any variation $\left\vert \delta
\psi\right\rangle $ with $\left\langle \delta\psi|\psi\right\rangle $ being
purely imaginary, Eq. (\ref{action4}) reduces to
\begin{equation}
\frac{d}{ds}\frac{\left\vert u_{\perp}\right\rangle }{\left\Vert \left\vert
u_{\perp}\right\rangle \right\Vert }-\left\langle \psi|u\right\rangle
\frac{\left\vert u_{\perp}\right\rangle }{\left\Vert \left\vert u_{\perp
}\right\rangle \right\Vert }=f\left(  s\right)  \left\vert \psi\right\rangle
\text{,} \label{GEq}%
\end{equation}
where $f\left(  s\right)  $ is an arbitrary real-valued function. Recalling
that $\left\langle u|\psi\right\rangle $ is purely imaginary, we can set
$\left\langle \psi|u\right\rangle \overset{\text{def}}{=}iA_{\psi}\left(
u\right)  $ with $A_{\psi}\left(  u\right)  \in%
\mathbb{R}
$. In this manner, the geodesic equation in Eq. (\ref{GEq}) becomes%
\begin{equation}
\left[  \frac{d}{ds}-iA_{\psi}\left(  u\right)  \right]  \frac{\left\vert
u_{\perp}\right\rangle }{\left\Vert \left\vert u_{\perp}\right\rangle
\right\Vert }=f\left(  s\right)  \left\vert \psi\right\rangle \text{.}
\label{georel}%
\end{equation}
At this point, we remark that\textbf{\ }geodesic paths can be formally
obtained by integrating the geodesic relation in Eq. (\ref{georel}). Since the
action functional $\mathcal{L}\left[  \gamma_{\mathrm{geo}}\right]  $ is
invariant under reparametrizations and gauge transformations however, it
follows that geodesic paths are reparametrization and gauge covariant
quantities. Specifically, we observe that by exploiting the gauge freedom,
$\gamma_{\mathrm{geo}}\left(  s\right)  $ is a geodesic path traced by a
vector state $\left\vert \psi\left(  s\right)  \right\rangle $ so that the
horizontal lift $\bar{\gamma}_{\mathrm{geo}}\left(  s\right)  $ of
$\gamma_{\mathrm{geo}}\left(  s\right)  $ is\textbf{\ }traced by the state
vector $\left\vert \psi_{\mathrm{h}}\left(  s\right)  \right\rangle $ with
$\left\langle \psi_{\mathrm{h}}\left(  s\right)  |\partial_{s}\psi
_{\mathrm{h}}\left(  s\right)  \right\rangle =0$. Therefore, by taking
advantage of the gauge freedom, we set $A_{\psi_{\mathrm{h}}}\left(  u\right)
\overset{\text{def}}{=}-i\left\langle \psi_{\mathrm{h}}|u\right\rangle =0$
with the consequence being that $\left\vert u_{\perp}\right\rangle =\left\vert
u\right\rangle $. Thus, Eq. (\ref{georel}) becomes,%
\begin{equation}
\frac{d}{ds}\frac{\left\vert u_{\perp}\right\rangle }{\left\Vert \left\vert
u_{\perp}\right\rangle \right\Vert }=f\left(  s\right)  \left\vert
\psi_{\mathrm{h}}\right\rangle \text{.} \label{bethere}%
\end{equation}
At this stage, by exploiting the parametrization covariance, we can choose a
convenient affine parametrization of the curve which is unique (modulo linear
inhomogeneous changes in $s$, $s\rightarrow\tilde{s}=as+b$ with $a$, $b\in%
\mathbb{R}
\backslash\left\{  0\right\}  $) such that $\left\Vert u\right\Vert $ is
constant along the curve. Therefore, upon\textbf{\ }setting $\left\langle
\dot{\psi}_{\mathrm{h}}\left(  s\right)  |\dot{\psi}_{\mathrm{h}}\left(
s\right)  \right\rangle =$ \textrm{constant}, $\left\langle \psi_{\mathrm{h}%
}\left(  s\right)  |\psi_{\mathrm{h}}\left(  s\right)  \right\rangle =1$, and
$\left\langle \psi_{\mathrm{h}}\left(  s\right)  |\dot{\psi}_{\mathrm{h}%
}\left(  s\right)  \right\rangle =$ $0$ for any $s$, we find that an
horizontal affinely parametrized geodesic in the space of unit rays satisfies
the equation%
\begin{equation}
\frac{d^{2}}{ds^{2}}\left\vert \psi_{\mathrm{h}}\left(  s\right)
\right\rangle +\left\langle \dot{\psi}_{\mathrm{h}}\left(  s\right)
|\dot{\psi}_{\mathrm{h}}\left(  s\right)  \right\rangle \left\vert
\psi_{\mathrm{h}}\left(  s\right)  \right\rangle =0\text{.} \label{sho}%
\end{equation}
The differential relation in Eq. (\ref{sho}) describes a simple harmonic
oscillator. Assuming $\left\langle \psi_{\mathrm{h}}\left(  0\right)
|\psi_{\mathrm{h}}\left(  0\right)  \right\rangle =1$, $\left\langle
\psi_{\mathrm{h}}\left(  0\right)  |\dot{\psi}_{\mathrm{h}}\left(  0\right)
\right\rangle =$ $0$, and $\left\langle \dot{\psi}_{\mathrm{h}}\left(
0\right)  |\dot{\psi}_{\mathrm{h}}\left(  0\right)  \right\rangle
=v_{\mathrm{H}}^{2}$, the general solution $\left\vert \psi_{\mathrm{h}%
}\left(  s\right)  \right\rangle $ of Eq. (\ref{sho}) becomes,%
\begin{equation}
\left\vert \psi_{\mathrm{h}}\left(  s\right)  \right\rangle =\cos\left(
v_{\mathrm{H}}s\right)  \left\vert \psi_{\mathrm{h}}\left(  0\right)
\right\rangle +\frac{\sin\left(  v_{\mathrm{H}}s\right)  }{v_{\mathrm{H}}%
}\left\vert \dot{\psi}_{\mathrm{h}}\left(  0\right)  \right\rangle \text{.}
\label{geostar}%
\end{equation}
The emergence of the horizontal, affinely parametrized geodesic path in Eq.
(\ref{geostar}) concludes our formal derivation.

As a final remark, we point out that we refer to the third subsection of this
Appendix for a brief physical note on the important concepts of dynamical and
geometric phases (see Eq. (\ref{raysfunctional})) in quantum mechanical evolutions.

\subsection{Nonhorizontal nonaffinely parametrized geodesics}

We recall that any two arbitrary non-orthogonal vectors $\left\vert \psi
_{A}\right\rangle $ and $\left\vert \psi_{B}\right\rangle $ in the unit ray
space can be connected by a geodesic arc that is generally non-horizontal and
non-affinely parametrized \cite{mukunda93},%
\begin{equation}
\left\vert \psi\left(  s\right)  \right\rangle =e^{i\beta\frac{s}{\theta}%
}\left[  \sin\left(  \theta-s\right)  \left\vert \psi_{A}\right\rangle
+e^{-i\beta}\sin\left(  s\right)  \left\vert \psi_{B}\right\rangle \right]
\text{,} \label{geonostar}%
\end{equation}
where $0\leq s\leq\theta$ with $0<\theta<\pi/2$ and $\left\langle \psi
_{A}|\psi_{B}\right\rangle =\left\vert \left\langle \psi_{A}|\psi
_{B}\right\rangle \right\vert e^{i\beta}$. Note that if $\left\langle \psi
_{A}|\psi_{B}\right\rangle =\delta_{AB}$, it is sufficient to take $\beta=0$
and $\theta=\pi/2$ in order to obtain a geodesic arc $\left\vert \psi\left(
s\right)  \right\rangle $ connecting the two orthogonal states $\left\vert
\psi_{A}\right\rangle $ and $\left\vert \psi_{B}\right\rangle $. The geodesic
in Eq. (\ref{geonostar}) is generally neither horizontal nor affinely
parametrized. The horizontality condition $\left\langle \psi\left(  s\right)
|\dot{\psi}\left(  s\right)  \right\rangle =$ $0$ is achieved via gauge
freedom, while the affine parametrization condition $\left\langle \dot{\psi
}\left(  s\right)  |\dot{\psi}\left(  s\right)  \right\rangle =$
\textrm{constant }is obtained via reparametrization freedom.

For example, let $\left\vert \psi\left(  t\right)  \right\rangle $ represent
the state vector of a quantum system that evolves according to the
Schr\"{o}dinger equation $i\hslash\partial_{t}\left\vert \psi\left(  t\right)
\right\rangle =\mathrm{H}\left\vert \psi\left(  t\right)  \right\rangle $. If
one exploits only the gauge freedom, then the horizontal vector $\left\vert
\psi\left(  t\right)  \right\rangle $ satisfies the equation%
\begin{equation}
\frac{d^{2}}{dt^{2}}\left\vert \psi\left(  t\right)  \right\rangle +\left[
v_{\mathrm{H}}\left(  t\right)  \right]  ^{2}\left\vert \psi\left(  t\right)
\right\rangle =0\text{,} \label{love}%
\end{equation}
where $v_{\mathrm{H}}\left(  t\right)  \overset{\text{def}}{=}\Delta E\left(
t\right)  /\hslash$ with $\Delta E\left(  t\right)  $ being the uncertainty in
the energy of the system and $t$ is the ordinary physical time parameter. To
verify that $v_{\mathrm{H}}\left(  t\right)  =\Delta E\left(  t\right)
/\hslash$, recall that $\left\vert u_{\perp}\right\rangle \overset{\text{def}%
}{=}\left\vert u\right\rangle -\left\langle \psi|u\right\rangle \left\vert
\psi\right\rangle $ and $\left\vert u\right\rangle \overset{\text{def}}%
{=}\left\vert \dot{\psi}\right\rangle =-\frac{i}{\hslash}\mathrm{H}\left\vert
\psi\right\rangle $. Then, we have%
\begin{align}
\left[  v_{\mathrm{H}}\left(  t\right)  \right]  ^{2}  &  =\left\langle
u_{\perp}|u_{\perp}\right\rangle \nonumber\\
&  =\left\langle u|u\right\rangle -\left\vert \left\langle u|\psi\right\rangle
\right\vert ^{2}\nonumber\\
&  =\left\langle \dot{\psi}|\dot{\psi}\right\rangle -\left\vert \left\langle
\dot{\psi}|\psi\right\rangle \right\vert ^{2}\nonumber\\
&  =\frac{1}{\hslash^{2}}\left[  \left\langle \psi|\mathrm{H}^{2}%
|\psi\right\rangle -\left\langle \psi|\mathrm{H}|\psi\right\rangle ^{2}\right]
\nonumber\\
&  =\frac{\left[  \Delta E\left(  t\right)  \right]  ^{2}}{\hslash^{2}%
}\text{,}%
\end{align}
that is, $v_{\mathrm{H}}\left(  t\right)  =\Delta E\left(  t\right)  /\hslash
$. In general, when a non-affine parameter is used to describe the geodesic
curve, the analytical integration of Eq. (\ref{love}) can be highly nontrivial.

\subsection{Dynamical and geometric phases in quantum evolutions}

From a physics standpoint, the concept of geometric phase (also known as
Berry's phase, \cite{berry84}) emerges when considering the adiabatic
evolution of a quantum mechanical system whose Hamiltonian \textrm{H} returns
to its original value and the state vector evolves as an eigenstate of the
Hamiltonian. As the Hamiltonian returns to its original value after a time
$t$, the system will return to its original state, apart from a phase factor
$e^{i\varphi_{\text{tot}}\left(  t\right)  }$. Specifically, we have%
\begin{equation}
\left\vert \psi\left(  t\right)  \right\rangle =e^{i\varphi_{\text{tot}%
}\left(  t\right)  }\left\vert \psi\left(  0\right)  \right\rangle \text{.}
\label{berry}%
\end{equation}
The phase factor $e^{i\varphi_{\text{tot}}\left(  t\right)  }$ in Eq.
(\ref{berry}) can be described in terms of a circuit-dependent component
$e^{i\varphi_{\text{geo}}\left(  t\right)  }$ and an usual dynamical component
$e^{i\varphi_{\text{dyn}}\left(  t\right)  }=e^{-\frac{i}{\hslash}Et}$ which
specifies the evolution of any stationary state. In terms of phases, we have
\begin{equation}
\varphi_{\text{geo}}\left(  t\right)  =\varphi_{\text{tot}}\left(  t\right)
-\varphi_{\text{dyn}}\left(  t\right)  \text{.} \label{geo}%
\end{equation}
The total phase is defined as,%
\begin{equation}
\varphi_{\text{tot}}\left(  t\right)  \overset{\text{def}}{=}\arg\left[
\left\langle \psi\left(  0\right)  |\psi\left(  t\right)  \right\rangle
\right]  \text{,} \label{tot}%
\end{equation}
where $\arg\left(  z\right)  \overset{\text{def}}{=}\tan^{-1}\left[
\operatorname{Im}\left(  z\right)  /\operatorname{Re}\left(  z\right)
\right]  $ and $z\in%
\mathbb{C}
$. The dynamical phase is defined for any evolution, cyclic or not, and can be
expressed in terms of the time integral of the expectation value of the
Hamiltonian \textrm{H},%
\begin{equation}
\varphi_{\text{dyn}}\left(  t\right)  \overset{\text{def}}{=}-\frac{1}%
{\hslash}\int_{0}^{t}\left\langle \psi\left(  t^{\prime}\right)
|\mathrm{H}\left(  t^{\prime}\right)  |\psi\left(  t^{\prime}\right)
\right\rangle dt^{\prime}\text{.} \label{dyn}%
\end{equation}
The quantity $\varphi_{\text{dyn}}\left(  t\right)  $ in Eq. (\ref{dyn})
encodes information about the duration of the evolution of the physical
system. Substituting Eqs. (\ref{dyn}) and (\ref{tot}) into Eq. (\ref{geo}),
the geometric phase becomes%
\begin{equation}
\varphi_{\text{geo}}\left(  t\right)  =\arg\left[  \left\langle \psi\left(
0\right)  |\psi\left(  t\right)  \right\rangle \right]  +\frac{1}{\hslash}%
\int_{0}^{t}\left\langle \psi\left(  t^{\prime}\right)  |\mathrm{H}\left(
t^{\prime}\right)  |\psi\left(  t^{\prime}\right)  \right\rangle dt^{\prime
}\text{.} \label{wilde}%
\end{equation}
The geometric phase $\varphi_{\text{geo}}\left(  t\right)  $ in Eq.
(\ref{wilde}) is the so-called Berry phase. It offers relevant information
about the geometry of the path of the quantum evolution viewed in the
projective Hilbert space of rays. Observe that using Schr\"{o}dinger's
evolution equation, we have%
\begin{align}
\operatorname{Im}\int_{0}^{t}\left\langle \psi\left(  t^{\prime}\right)
|\dot{\psi}\left(  t^{\prime}\right)  \right\rangle dt^{\prime}  &
=\operatorname{Im}\left(  -\frac{i}{\hslash}\int_{0}^{t}\left\langle
\psi\left(  t^{\prime}\right)  |\mathrm{H}\left(  t^{\prime}\right)
|\psi\left(  t^{\prime}\right)  \right\rangle dt^{\prime}\right) \nonumber\\
&  =-\frac{1}{\hslash}\int_{0}^{t}\left\langle \psi\left(  t^{\prime}\right)
|\mathrm{H}\left(  t^{\prime}\right)  |\psi\left(  t^{\prime}\right)
\right\rangle dt^{\prime}\nonumber\\
&  =\varphi_{\text{dyn}}\left(  t\right)  \text{.} \label{wilde1}%
\end{align}
Therefore, identifying the corresponding quantities together with employing
Eqs. (\ref{wilde1}) and (\ref{wilde}), we can recover Eq.
(\ref{raysfunctional}). For further details on phase changes in cyclic and
noncyclic quantum evolutions, we refer to Ref. \cite{anandan87} and Ref.
\cite{pati95}, respectively. Finally, for a description of Berry's phase in
terms of natural geometric structures or in terms of the fiber bundle
language, we refer to Ref. \cite{page87} and Ref. \cite{bohm91}, respectively.

\section{Affinely v.s. nonaffinely parametrized geodesic paths}

In this Appendix, we present some technical details on the possibility of
using a non-affine parameter for describing the geodesic motion on a curved
spacetime manifold. This technicality was mentioned in Section IV when
discussing our gravitational problem.

In the framework of General Relativity on curved manifolds
\cite{defelice,stewart91}, an affinely parametrized geodesic $\gamma\left(
\tau\right)  $ is a curve whose tangent vector $\dot{\gamma}$ is everywhere
non-zero and is parallely propagated. Therefore, the covariant derivative of
the vector field $\dot{\gamma}$ along $\gamma$ is zero, that is $D_{\tau}%
\dot{\gamma}=0$. As a consequence, $\left\Vert \dot{\gamma}\right\Vert
=$\textrm{constant} since%
\begin{equation}
\partial_{\tau}\left\Vert \dot{\gamma}\right\Vert ^{2}\overset{\text{def}}%
{=}\left\langle D_{\tau}\dot{\gamma}\text{, }\dot{\gamma}\right\rangle
+\left\langle \dot{\gamma}\text{, }D_{\tau}\dot{\gamma}\right\rangle
=0\text{.}%
\end{equation}
In particular, the differential equation for an affine geodesic $\gamma\left(
\tau\right)  $, with $\tau$ being the affine parameter along the curve is
given by,%
\begin{equation}
\frac{d^{2}\gamma^{a}}{d\tau^{2}}+\Gamma_{bc}^{a}\frac{d\gamma^{b}}{d\tau
}\frac{d\gamma^{c}}{d\tau}=0\text{,} \label{age}%
\end{equation}
with $\Gamma_{bc}^{a}$ being the usual Christoffel connection coefficients of
the second kind. The standard form of Eq. (\ref{age}) is preserved if and only
if we replace $\tau$ with $\tilde{\tau}=\tilde{\tau}\left(  \tau\right)
\overset{\text{def}}{=}A\tau+B$ with $A$, $B$ being constants. Stated
otherwise, an affine parameter is defined up to a change of scale ($A\neq0$)
and origin ($B\neq0$). More generally, a geodesic curve $\gamma\left(
s\right)  $ is a curve whose tangent vector $\dot{\gamma}$ is everywhere
non-zero\textbf{\ }and only needs to be proportional to a parallely propagated
vector. In this case, the non-affinely parametrized geodesic equation becomes,%
\begin{equation}
\frac{d^{2}\gamma^{a}}{ds^{2}}+\Gamma_{bc}^{a}\frac{d\gamma^{b}}{ds}%
\frac{d\gamma^{c}}{ds}=g\left(  s\right)  \frac{d\gamma^{a}}{ds}\text{.}
\label{nage}%
\end{equation}
The quantity $g\left(  s\right)  $ in Eq. (\ref{nage}) is defined as $g\left(
s\right)  \overset{\text{def}}{=}\frac{d}{ds}\ln\left[  \lambda^{-1}\left(
s\right)  \right]  $ with $\lambda\left(  s\right)  >0$ being a differentiable
function such that \cite{defelice},%
\begin{equation}
\left[  \Gamma\left(  0\text{, }s\text{; }\gamma\right)  \right]  _{a}%
^{b}\cdot\left[  \frac{d\gamma^{a}}{ds}\left(  0\right)  \right]
=\lambda\left(  s\right)  \left[  \frac{d\gamma^{b}}{ds}\left(  s\right)
\right]  \text{,}%
\end{equation}
with $\Gamma\left(  0\text{, }s\text{; }\gamma\right)  $ denoting the
so-called connector map. As a final remark, we point out that it can be shown
that Eq. (\ref{nage}) reduces to Eq. (\ref{age}) by performing a suitable
change of variables,%
\begin{equation}
s\rightarrow\tau:s=\sigma\left(  \tau\right)  \text{, with }\frac{d\sigma
}{d\tau}\overset{\text{def}}{=}\lambda\left(  \sigma\left(  \tau\right)
\right)  \text{.}%
\end{equation}
We leave this simple verification as an exercise for the interested reader.

\section{Geometrization of Newtonian Mechanics}

In this Appendix, we present a straightforward derivation of Newton's equation
of motion in Eq. (\ref{tsho}) starting from Eq. (\ref{geoshit}). This
technicality was mentioned in Section IV when discussing our thermodynamical problem.

Recall that in a local coordinate system the equation of an affinely
parametrized geodesic is given by%
\begin{equation}
\frac{d^{2}q^{i}}{ds^{2}}+\Gamma_{jk}^{i}\frac{dq^{j}}{ds}\frac{dq^{k}}%
{ds}=0\text{.} \label{jojo}%
\end{equation}
The Christoffel symbols $\Gamma_{jk}^{i}$ in Eq. (\ref{jojo}) are defined as,%
\begin{equation}
\Gamma_{jk}^{i}\overset{\text{def}}{=}\frac{1}{2}g^{im}\left(  \partial
_{j}g_{km}+\partial_{k}g_{mj}-\partial_{m}g_{jk}\right)  \text{,}
\label{Chris}%
\end{equation}
where $g_{ij}\overset{\text{def}}{=}2\left[  E-V\left(  q\right)  \right]
\delta_{ij}$. Using Eq. (\ref{Chris}) together with the expression of the
Jacobi metric, Eq. (\ref{jojo}) becomes%
\begin{align}
0  &  =\frac{d^{2}q^{i}}{ds^{2}}+\Gamma_{jk}^{i}\frac{dq^{j}}{ds}\frac{dq^{k}%
}{ds}\nonumber\\
&  =\frac{d^{2}q^{i}}{ds^{2}}+\frac{1}{2}g^{im}\left(  \partial_{j}%
g_{km}+\partial_{k}g_{mj}-\partial_{m}g_{jk}\right)  \frac{dq^{j}}{ds}%
\frac{dq^{k}}{ds}\nonumber\\
&  =\frac{d^{2}q^{i}}{ds^{2}}+\frac{1}{2}g^{im}\partial_{j}g_{km}\frac{dq^{j}%
}{ds}\frac{dq^{k}}{ds}+\frac{1}{2}g^{im}\partial_{k}g_{mj}\frac{dq^{j}}%
{ds}\frac{dq^{k}}{ds}-\frac{1}{2}g^{im}\partial_{m}g_{jk}\frac{dq^{j}}%
{ds}\frac{dq^{k}}{ds}\nonumber\\
&  =\frac{d^{2}q^{i}}{ds^{2}}+\frac{1}{2}\frac{1}{\left[  E-V\left(  q\right)
\right]  }\delta^{im}\frac{\partial\left\{  \left[  E-V\left(  q\right)
\right]  \delta_{km}\right\}  }{\partial q^{j}}\frac{dq^{j}}{ds}\frac{dq^{k}%
}{ds}+\nonumber\\
&  +\frac{1}{2}\frac{1}{\left[  E-V\left(  q\right)  \right]  }\delta
^{im}\frac{\partial\left\{  \left[  E-V\left(  q\right)  \right]  \delta
_{mj}\right\}  }{\partial q^{k}}\frac{dq^{j}}{ds}\frac{dq^{k}}{ds}+\nonumber\\
&  -\frac{1}{2}\frac{1}{\left[  E-V\left(  q\right)  \right]  }\delta
^{im}\frac{\partial\left\{  \left[  E-V\left(  q\right)  \right]  \delta
_{jk}\right\}  }{\partial q^{m}}\frac{dq^{j}}{ds}\frac{dq^{k}}{ds}\nonumber\\
&  =\frac{d^{2}q^{i}}{ds^{2}}+\frac{1}{2}\frac{1}{\left[  E-V\left(  q\right)
\right]  }\frac{\partial\left[  E-V\left(  q\right)  \right]  }{\partial
q^{j}}\frac{dq^{j}}{ds}\frac{dq^{i}}{ds}+\frac{1}{2}\frac{1}{\left[
E-V\left(  q\right)  \right]  }\frac{\partial\left[  E-V\left(  q\right)
\right]  }{\partial q^{k}}\frac{dq^{i}}{ds}\frac{dq^{k}}{ds}+\nonumber\\
&  -\frac{1}{2}\frac{1}{\left[  E-V\left(  q\right)  \right]  }\frac
{\partial\left[  E-V\left(  q\right)  \right]  }{\partial q_{i}}\left(
\frac{dq^{j}}{ds}\right)  ^{2}\nonumber\\
&  =\frac{d^{2}q^{i}}{ds^{2}}+\frac{1}{2}\frac{1}{\left[  E-V\left(  q\right)
\right]  }\left[  2\frac{\partial\left[  E-V\left(  q\right)  \right]
}{\partial q^{j}}\frac{dq^{j}}{ds}\frac{dq^{i}}{ds}-g^{ij}\frac{\partial
\left[  E-V\left(  q\right)  \right]  }{\partial q^{j}}g_{km}\frac{dq^{k}}%
{ds}\frac{dq^{m}}{ds}\right]  \text{,}%
\end{align}
that is,%
\begin{equation}
\frac{d^{2}q^{i}}{ds^{2}}+\frac{1}{2}\frac{1}{\left[  E-V\left(  q\right)
\right]  }\left[  2\frac{\partial\left[  E-V\left(  q\right)  \right]
}{\partial q^{j}}\frac{dq^{j}}{ds}\frac{dq^{i}}{ds}-g^{ij}\frac{\partial
\left[  E-V\left(  q\right)  \right]  }{\partial q^{j}}g_{km}\frac{dq^{k}}%
{ds}\frac{dq^{m}}{ds}\right]  =0\text{.} \label{she}%
\end{equation}
Next, recalling that $ds^{2}\overset{\text{def}}{=}g_{ij}dq^{i}dq^{j}=4\left[
E-V\left(  q\right)  \right]  ^{2}dt^{2}$, Eq. (\ref{she}) becomes%
\begin{equation}
\frac{d^{2}q^{i}}{dt^{2}}+\frac{\partial V\left(  q\right)  }{\partial q_{i}%
}=0\text{,}%
\end{equation}
that is,%
\begin{equation}
\frac{d^{2}q^{i}}{dt^{2}}=-\frac{\partial V\left(  q\right)  }{\partial q_{i}%
}\text{.}%
\end{equation}
Finally, observing that $n_{\text{\textrm{dof}}}=1$, $q=x$, and the potential
$V\left(  q\right)  $ in Eq. (\ref{Jacobi}) reduces to the harmonic potential
$(1/2)mv_{\text{th}}^{2}x^{2}$ with $v_{\text{th}}$ defined in Eq.
(\ref{omega}), we get Eq. (\ref{tsho}). For further details on the Riemannian
geometrization of Newtonian mechanics, we refer to Refs.
\cite{casetti00,pettini07}.

\begin{thebibliography}{99}                                                                                               %


\bibitem {frieden98}B. R. Frieden, \emph{Physics from Fisher Information},
Cambridge University Press, New York (1998).

\bibitem {cover06}T. M. Cover and J. A. Thomas, \emph{Elements of Information
Theory}, John Wiley \& Sons, New York (2006).

\bibitem {Felice18}D. Felice, C. Cafaro and S. Mancini, \emph{Information
geometric methods for complexity}, {Chaos} \textbf{28}, {032101} (2018).

\bibitem {carlopre}C. Cafaro and P. M.\ Alsing, \emph{Decrease of Fisher
information and the information geometry of evolution equations for quantum
mechanical probability amplitudes}, Phys. Rev. \textbf{E97}, 042110 (2018).

\bibitem {villani00}C. Villani, \emph{Decrease of the Fisher information for
the Landau equation with Maxwellian molecules}, Math. Mod. Meth. Appl. Sci.
\textbf{10}, 153 (2000).

\bibitem {alvarez00}J. J. Alvarez and C. Gomez, \emph{A comment on Fisher
information and quantum algorithms}, arXiv:quant-ph/9910115 (2000).

\bibitem {wadati01}A. Miyake and M. Wadati, \emph{Geometric strategy for the
optimal quantum search}, Phys. Rev. \textbf{A64}, 042317 (2001).

\bibitem {carlophysica}C. Cafaro, \emph{Geometric algebra and information
geometry for quantum computational software}, Physica \textbf{A470}, 154 (2017).

\bibitem {carloIJQI}C. Cafaro and P. M. Alsing, \emph{Continuous-time quantum
search and time-dependent two-level quantum systems}, Int. J. Quantum
Information \textbf{17}, 1950025 (2019).

\bibitem {carloPS}C. Cafaro\textbf{\ }and P. M. Alsing, \emph{Theoretical
analysis of a nearly optimal analog quantum search}, Physica Scripta
\textbf{94}, 085103 (2019).

\bibitem {gassner20}S. Gassner, C. Cafaro, and S. Capozziello,
\emph{Transition probabilities in generalized quantum search Hamiltonian
evolutions}, Int. Journal of Geometric Methods in Modern Physics \textbf{17},
2050006 (2020).

\bibitem {carlopre20}C. Cafaro and P. M.\ Alsing, \emph{Information geometry
aspects of minimum entropy production paths from quantum mechanical
evolutions}, Physical Review \textbf{E101}, 022110 (2020).

\bibitem {quantum20}C. Cafaro, S. Gassner, and P. M. Alsing, \emph{Information
geometric perspective on off-resonance effects in driven two-level quantum
systems}, Quantum Reports \textbf{2, }166 (2020).

\bibitem {mukunda93}N. Mukunda and R. Simon, \emph{Quantum kinematic approach
to the geometric phase.I. General formalism}, Annals of Physics \textbf{228},
205 (1993).

\bibitem {pati94}A. K. Pati, \emph{On phases and length of curves in a cyclic
quantum evolution}, Pramana-J. Phys. \textbf{42}, 455 (1994).

\bibitem {pati95}A. K. Pati, \emph{Geometric aspects of noncyclic quantum
evolutions}, Phys. Rev. \textbf{A52}, 2576 (1995).

\bibitem {mancioA}C. Cafaro and S. Mancini, \emph{An information geometric
viewpoint of algorithms in quantum computing}, AIP Conf. Proc. \textbf{1443},
374 (2012).

\bibitem {mancioB}C. Cafaro and S. Mancini, \emph{On Grover's search algorithm
from a quantum information geometry viewpoint}, Physica \textbf{A391}, 1610 (2012).

\bibitem {Ciaglia18}F. M. Ciaglia, F. Di Cosmo, D. Felice, S. Mancini, G.
Marmo and J.M. P\'{e}rez-Pardo, \emph{Aspects of geodesical motion with
Fisher-Rao metric: Classical and Quantum}, Open Systems and Information
Dynamics \textbf{25}, 1850005 (2018).

\bibitem {anandan90}J. Anandan and Y. Aharonov, \emph{Geometry of quantum
evolution}, Phys. Rev. Lett. \textbf{65}, 1697 (1990).

\bibitem {grover}L. K. Grover, \emph{Quantum\ mechanics helps in searching for
a needle in a haystack}, Phys. Rev. Lett. \textbf{79}, 325 (1997).

\bibitem {zatzkis59}H. Zatzikis, \emph{Model of a linear harmonic oscillator
in the general theory of relativity}, Phys. Rev. \textbf{114}, 1645 (1959).

\bibitem {ruchhardt29}E. R\"{u}chhardt, \emph{Eine einfache methode zur
bestimmung von }$C_{P}/C_{V}$, Physikalische Zeitschrift \textbf{30}, 58 (1929).

\bibitem {nielsen}M. A. Nielsen and I. L. Chuang, \emph{Quantum Computation
and Quantum Information}, Cambridge University Press (2000).

\bibitem {provost80}J. P. Provost and G. Vallee, \emph{Riemannian structure on
manifolds of quantum states}, Commun. Math. Phys. \textbf{76}, 289 (1980).

\bibitem {braunstein94}S. L. Braunstein and C. M. Caves, \emph{Statistical
distance and the geometry of quantum states}, Phys. Rev. Lett. \textbf{72},
3439 (1994).

\bibitem {wootters81}W. K. Wootters, \emph{Statistical distance and Hilbert
space}, Phys. Rev. \textbf{D23}, 357 (1981).

\bibitem {Marmo}P. Facchi, R. Kulkarni, V. I. Man'ko, G. Marmo, E.C.G.
Sudarshan , and F. Ventriglia, \emph{Classical and quantum Fisher information
in the geometrical formulation of quantum mechanics}, Physics Letters
\textbf{A374}, 4801 (2010).

\bibitem {farhi00}E. Farhi, J. Goldstone, S. Gutmann, and M. Sipser,
\emph{Quantum computation by adiabatic evolution}, arXiv:quant-ph/0001106 (2000).

\bibitem {zanardi09}A. T. Rezakhani, W.-J. Kuo, A. Hamma, D. A. Lidar, and P.
Zanardi, \emph{Quantum adiabatic brachistochrone}, Phys. Rev. Lett.
\textbf{103}, 080502 (2009).

\bibitem {roland02}J. Roland and N. J. Cerf, \emph{Quantum search by local
adiabatic evolution}, Phys. Rev. \textbf{A65}, 042308 (2002).

\bibitem {zanardi10}A. T. Rezakhani, D. F. Abasto, D. A. Lidar, and P.
Zanardi, \emph{Intrinsic geometry of quantum adiabatic evolution and quantum
phase transitions}, Phys. Rev.\ \textbf{A82}, 012321 (2010).

\bibitem {defelice}F. De Felice and C. J. S. Clarke, \emph{Relativity on
Curved Manifolds}, Cambridge University Press (1990).

\bibitem {laue21}M. von Laue, \emph{Die Relativit\"{a}tstheorie}, Vol. 2, F.
Vieweg und Sohn, Braunschweig (1921).

\bibitem {fermi94}E. Fermi, \emph{Termodinamica}, Bollati Boringhieri (1994).

\bibitem {zemansky97}M. W. Zemansky and R. H. Dittman, \emph{Heat and
Thermodynamics}, The McGraw-Hill Companies, Inc. (1997).

\bibitem {casetti00}L. Casetti, M. Pettini, and E. G. D. Cohen,
\emph{Geometric approach to Hamiltonian dynamics and statistical mechanics},
Phys. Rep. \textbf{337}, 237 (2000).

\bibitem {caticha07}A. Caticha and C. Cafaro, \emph{From information geometry
to Newtonian dynamics}, AIP Conf. Proc. \textbf{954}, 165 (2007).

\bibitem {cafaro09}C. Cafaro, \emph{Works on an information geometrodynamical
approach to chaos}, Chaos, Solitons and Fractals \textbf{41}, 886 (2009).

\bibitem {padmanabhan}T. Padmanabhan, \emph{Gravitation: Foundations and
Frontiers}, Cambridge University Press (2010).

\bibitem {luongo16}P. K. S. Dunsby, O. Luongo, and L. Reverberi, \emph{Dark
energy and dark matter from an additional adiabatic fluid}, Phys. Rev.
\textbf{D94}, 083525 (2016).

\bibitem {goldstein}H. Goldstein, \emph{Meccanica Classica}, Zanichelli S. p.
A. (1971).

\bibitem {jaynes85}E. T. Jaynes, \emph{Macroscopic prediction}, in Complex
Systems- Operational Approaches in Neurobiology, Physics, and Computers, H.
Haken, Ed.; Springer-Verlag, Berlin, pp. 254 (1985).

\bibitem {cafaro16}C. Cafaro and S. A. Ali, \emph{Maximum caliber inference
and the stochastic Ising model}\textbf{, }Phys. Rev. \textbf{E94}, 052145 (2016).

\bibitem {karol}I. Bengtsson and K. Zyczkowski, \emph{Geometry of Quantum
States}, Cambridge University Press (2006).

\bibitem {brody96}D. C. Brody and L. P. Hughston, \emph{Geometry of quantum
statistical inference}, Phys. Rev. Lett. \textbf{77}, 2851 (1996).

\bibitem {pezze09}L. Pezz\'{e} and A.\ Smerzi, \emph{Entanglement, nonlinear
dynamics, and the Heisenberg limit}, Phys. Rev. Lett. \textbf{102}, 100401 (2009).

\bibitem {taddei13}M. M. Taddei, B. M. Escher, L. Davidovich, and R. L. de
Matos Filho, \emph{Quantum speed limit for physical processes}, Phys. Rev.
Lett. \textbf{110}, 050402 (2013).

\bibitem {boixo07}S. Boixo, S. T. Flammia, C. M. Caves, and JM Geremia,
\emph{Generalized limits for single-parameter quantum estimation}, Phys. Rev.
Lett. \textbf{98}, 090401 (2007).

\bibitem {vecchi83}S. Kirkpatrick, C. D. Gelatt Jr., and M. P. Vecchi,
\emph{Optimization by simulated annealing}, Science\textbf{\ 220}, 671 (1983).

\bibitem {berry84}M. V. Berry, \emph{Quantal phase factors accompanying
adiabatic changes}, Proceedings of the Royal Society London, Ser.
\textbf{A392}, 45 (1984).

\bibitem {anandan87}Y. Aharonov and J. Anandan, \emph{Phase change during a
cyclic quantum evolution}, Phys. Rev. Lett. \textbf{58}, 1593 (1987).

\bibitem {page87}D. N. Page, \emph{Geometrical description of Berry's phase},
Phys. Rev. \textbf{A36}, 3479 (1987).

\bibitem {bohm91}A. Bohm, L. J. Boya, and B. Kendrick, \emph{Derivation of the
geometrical phase}, Phys. Rev. \textbf{A43}, 1206 (1991).

\bibitem {stewart91}J. Stewart, \emph{Advanced General Relativity}, Cambridge
University Press (1991).

\bibitem {pettini07}M. Pettini, \emph{Geometry and Topology in Hamiltonian
Dynamics and Statistical Mechanics}, Springer-Verlag New York Inc. (2007).
\end{thebibliography}
\end{document}